\documentclass{JHEP3} 

\usepackage{epsfig}
\usepackage{amsfonts}
\usepackage{amssymb}

\newcommand{\V}{{\mathbb{V}}}
\newcommand{\PP}{{\mathbb{P}}}

\newcommand{\x}{\vec{x}}
\newcommand{\y}{\vec{y}}
\newcommand{\kk}{\vec{k}}

\newcommand{\be}{\begin{equation}}
\newcommand{\ee}{\end{equation}}

\newcommand\fverb{\setbox\pippobox=\hbox\bgroup\verb}
\newcommand\fverbdo{\egroup\medskip\noindent%
                        \fbox{\unhbox\pippobox}\ }
\newcommand\fverbit{\egroup\item[\fbox{\unhbox\pippobox}]}
\newbox\pippobox

\title{Classical Solutions of the TEK Model and Noncommutative Instantons
in Two Dimensions}

\author{ Luca Griguolo \\
Dipartimento di  Fisica, Universit\`a  di Parma,
INFN-Gruppo Collegato di Parma\\
Parco Area delle Scienze 7/A, 43100 Parma, Italy\\
E-mail: \email{griguolo@fis.unipr.it}}
\author{Domenico Seminara\\
Dipartimento di Fisica, Polo Scientifico Universit\`a di Firenze,
INFN Sezione di Firenze\\
Via  G. Sansone 1, 50019 Sesto Fiorentino, Italy\\
Email: \email{seminara@fi.infn.it}
}

\received{\today}               

\accepted{\today}               
\preprint{UPRF-2003-09}      

\date{data}
\abstract{The twisted Eguchi-Kawai (TEK) model provides a non-perturbative
definition of noncommutative Yang-Mills theory: the
continuum limit is approached at large $N$ by performing suitable
double scaling limits, in which non-planar contributions are no
longer suppressed. We consider here the two-dimensional case,
trying to recover within this framework the exact results recently
obtained by means of Morita equivalence. We present a rather
explicit construction of classical gauge theories on
noncommutative toroidal lattice for general topological charges.
After discussing the limiting procedures to recover the theory on
the noncommutative torus and on the noncommutative plane, we focus
our attention on the classical solutions of the related TEK
models. We solve the equations of motion and we find the
configurations having finite action in the relevant double scaling
limits. They can be explicitly described in terms of twist-eaters
and they exactly correspond to the instanton solutions that are
seen to dominate the partition function on the
noncommutative torus. Fluxons on the noncommutative plane are
recovered as well. We also discuss how the highly non-trivial
structure of the exact partition function can emerge from a direct
matrix model computation. The quantum consistency of the TEK
formulation is eventually checked by computing Wilson loops in a
particular limit.}

\keywords{Noncommutative Gauge Theories, Matrix Models, Large-N limit}

\begin{document}

\section{Introduction}
\noindent
The possibility to embed consistently a noncommutative field theory
into a string theory \cite{cds,sw} has stimulated in the last years a large amount
of studies, trying to understand classical and quantum noncommutative
dynamics both at perturbative and non-perturbative level (see \cite{dn,s}
for a review). In particular quantum field theories on
noncommutative spacetimes represent a framework in which to study D-branes physics retaining part of the
non-locality inherent in string
theory. From a purely field theoretical point of view, instead, they appear as a
highly non-trivial non-local deformation of conventional quantum field
theory, presenting a large variety of new phenomena not completely
understood even in the basic cases: at perturbative level the UV/IR
mixing \cite{msr,suski} complicates the renormalization program and it may
 produce tachyonic instabilities \cite{espe}. In both cases an
intriguing interplay between perturbative and non-perturbative effects
seems to conspire in order to recover a consistent physical picture: in
\cite{gp} it was shown that resummation of perturbation theory is
mandatory to have a sensible infrared limit in four-dimensional scalar
theories while the results of \cite{adi} seems to indicate that the
presence of new (extended) degrees of freedom may overcome the
instability problem. The possibility of transitions to new "exotic" phases has also
been put forward \cite{gubso}.

\noindent
On the other hand the analysis presented in \cite{sm}
clearly suggests that UV/IR mixing is not a perturbative artefact, being
intimately related to the non-perturbative structure of
noncommutative gauge theories. Lattice regularization with periodic boundary conditions
is in fact equivalent, in this case, to a well-known unitary (multi)matrix model,
the twisted Eguchi-Kawai (TEK) model \cite{gonzo}: in the usual large-$N$ limit it is
believed to reproduce Yang-Mills theory in the 't Hooft regime (see \cite{das} for a
review on the subject). The novelty of its noncommutative incarnation is that to
reach the continuum limit one has to perform a ${\it double\,scaling}$ limit,
in which the dimensionful lattice spacing scales to zero with a precise power of $N$:
non-planar contributions are no longer suppressed and carry the relevant
physics of the noncommutative theory.

\noindent The possibility to have an explicit non-perturbative
formulation of noncommutative gauge theories through matrix models
is not unexpected of course: noncommutative Yang-Mills theory was
directly obtained from the large-$N$ limit of the $IIB$ matrix model
\cite{aoki}. The spacetime dependence emerged from expanding
around a classical vacuum, but initially it is hidden in the
infinitely many degrees of freedom of the (large) matrices: on the
other hand the original appearance of noncommutative geometry from
string theory was based on the very same mechanism \cite{cds}.

\noindent
A related and complementary approach to non-perturbative
physics of noncommutative gauge theories was
advocated in \cite{luis,gura2,gsv}, by using Morita equivalence
\cite{schwarz}. The basic idea is to start from the theory
compactified on a rational noncommutative torus $T^D$ (by rational
we mean  that the entries on the antisymmetric matrix
$\theta_{\mu\nu}$ describing the noncommutative star-product are
rational numbers when rescaled with the torus radii) and to reach
noncommutative $\mathbb{R}^D$ by a suitable decompactification limit. By
Morita equivalence the theory on a rational noncommutative torus
is equivalent to some commutative theory with 't Hooft fluxes. Then the
infinite space is attained by performing a large-$N$ limit that simultaneously
shrinks the size of the (commutative)
torus to zero with a specific powers of $N$. Again a
double scaling limit is required, confirming the non-perturbative
nature of UV/IR mixing. Particularly explicit results have been
obtained in the simplified context of two-dimensional
noncommutative gauge theories. In \cite{gsv}, employing the above strategy
and the known solution of the Morita equivalent theory \cite{mig},  we derived
 the partition function and  the correlators of
two Wilson lines on the noncommutative plane. A remarkable fact was
that the classical  configurations,
naturally dominating this limit, are in
correspondence with the noncommutative fluxons found
in \cite{polly,andy,gn2}. This result  also establishes  an unexpected relation
between the double scaling limit and classical solutions of the (noncommutative)
equations of motion.  In ordinary two-dimensional Yang-Mills theory, this connection is
instead absent: in fact no classical configuration survives the usual large-$N$ limit.

\noindent
Later in ref. \cite{pasz} a localization theorem for Yang-Mills theory on the noncommutative two-torus
was proven, generalizing the Witten's result \cite{witten} on the exactness of semiclassical
approximation for familiar two-dimensional gauge theories on Riemann surfaces. This beautiful result led
the authors to propose an interesting formula for the
exact partition function on the noncommutative torus, consistent with ours in the decompactification
limit. They were able to write the partition function, in the rational case, as an expansion around the
critical points of Yang-Mills theory on a particular projective module, exploiting the mentioned
equivalence with the parent commutative theory and the Poisson-resummed formulae presented in
\cite{grig}. Assuming smoothness on the noncommutative parameter, they obtained an intriguing
generalization to the irrational case: an important consistency check with the localization formula
was the peculiar form of the contribution of the quantum fluctuations, reflecting the singularity
structure of the moduli space of constant curvature connections \cite{cr}. Recently the same authors
extended their analysis to compute Wilson line correlation functions \cite{pasz2} and they
offered some evidences for relating the noncommutative theory to ordinary (commutative) generalized YM$_2$
\cite{dls}.

\noindent
The derivation of the above results (both ours and the ones contained in \cite{pasz,pasz2}), although giving
useful insights on the non-perturbative structure of quantum noncommutative theories, may be considered,
in some sense, unsatisfactory. In fact it heavily relies on the equivalence between theories with
${\it rational}$ noncommutative parameter and ordinary Yang-Mills theories and it takes advantage of
the exact solution of the latter in two dimensions (solution that is not available in higher dimensions of
course). On the other hand, as we mentioned before, a general non-perturbative formulation based on the
double scaling limit of TEK model exists and can be used in any dimension: it would be nice
to recover the results of \cite{gsv,pasz,pasz2} starting from this very fundamental definition of
noncommutative Yang-Mills theory by exploiting familiar matrix model/lattice techniques. Incidentally
TEK model in two dimensions was the subject of intense studies in the eighties \cite{molti},
in order to better understand its claimed equivalence with conventional Wilson approach to lattice gauge theories
at large $N$. At the time, of course, no interest in searching a non-trivial double scaling limit has been raised,
although later similar investigations on two-dimensional Wilson theory \cite{peri} were triggered by
matrix model approach to $2D$ quantum gravity \cite{matri}. Numerical studies \cite{prof,nisj} have been recently
performed, instead, to capture a double scaling limit in $2D$ TEK models, motivated by noncommutative
quantum field theory: those results exhibit significant deviations from the usual behaviors.

\noindent
In this paper we try to perform an analytical approach to the same problem: in particular our computations
should be considered as a first step in trying to recover the results presented in \cite{gsv} and
\cite{pasz,pasz2} starting from TEK models. We point out here a basic difference between the compact and
non-compact case: to recover the partition function on the noncommutative torus, one should have in fact to resort
to the ${\it constrained}$ TEK model proposed in \cite{sm}. There a peculiar double scaling limit was claimed
to reproduce the theory at finite area, the dimension of the matrices and the structure of the twists
being determined by Diophantine equations. The theory on the noncommutative plane is instead reached by
a different double scaling limit and without resorting to any constraint. The compact case is definitely
subtler: in particular it should account for non-trivial topological charges. The general description
of the two-dimensional case requires, in fact, the presence of two different integers, usually denoted as
$(p,q)$, classifying the inequivalent projective modules on a fixed noncommutative torus \cite{cr,cori}.
In order to perform concrete computations, an explicit parametrization of constrained TEK variables
and an efficient procedure to implement the double scaling limit are welcome. We have therefore
decided to reconsider the TEK formulation of gauge theories on the noncommutative two-dimensional
torus, starting from a slightly different point of view with respect to the approach of \cite{sm}.
The case of the noncommutative plane is easily recovered within the same framework. After, we have focused our
attention on the structure of the classical solutions of the model and on their behavior under double
scaling limits. The main result is that we are able to reproduce, within this framework, the whole tower
of instanton solutions on the $(p,q)$ projective module \cite{cr}, on which the partition function
has been shown to be localized \cite{pasz}. The fluxon configurations emerging on the noncommutative
plane \cite{gsv} are as
obtained as well, changing accordingly the double scaling limit.

\noindent
The plan of the paper is the following: In Sect. 2 we start by considering the structure of toroidal
noncommutative lattices and using reducible representations of Weyl-'t Hooft algebra we are able to construct
derivations endowed with an arbitrary constant curvature. They naturally depends on a pair of integers that
will be interpreted as the topological charges classifying the projective modules in the continuum limit.
We define the gauge theory
through a one-plaquette Wilson action, obtaining a related family of TEK models: the dimension of the matrices
and the structure of the twists nicely encode the topological content of the discretized modules. The equivalence
of our unconstrained systems with the constrained ones proposed in reference \cite{sm} is then carefully discussed.
In Sect. 3 we analyze the continuum limits and we derive explicitly the relevant scalings.
Sect. 4 is devoted to the solutions of the classical equations of motion of the two-dimensional twisted Eguchi-Kawai model,
showing that they fall into two different families, distinguished by their matrix structure. In Sect. 5 we describe
their double scaling behaviors: only one family is important here, the other having infinite action in both limits.
We find all the solutions having finite action and we discuss their equivalence with the
instanton solutions on the noncommutative torus and with the fluxons on the noncommutative plane.
The possibility to compute the partition function directly from the TEK formulation is discussed in Sect. 6: we remark
the differences with the (exactly solvable) commutative case and we outline a possible strategy. As a consistency
check of the TEK approach, we perform the computation, by matrix model technique, of the quantum average of a
Wilson loop of vanishing area but infinitely winding. In this limit the calculation can be done exactly
and we recover the known result of the noncommutative plane. In Sect. 7 we draw our conclusions and we present
the possible extensions of this work. In Appendix A we show that Morita equivalent theories are described, in
our formalism, by the very same matrix model. Appendix B is instead devoted to the explicit construction of the
Wilson lattice action associated to our TEK models by means of a discretized version of the familiar star-product.

\section{Noncommutative gauge theories in two dimensions and TEK models}
\noindent
Large-$N$ reduced models combine some powerful approaches to quantum field theories at non-perturbative
level: the $1/N$-expansion, the lattice approximation to space-time, the loop equations and the
matrix model techniques. The basic idea, dating back to Eguchi and Kawai \cite{ek}, is that standard
$U(N)$ and £$SU(N)$ lattice gauge theories may be equivalent to their reduction to one plaquette
in the large-$N$ limit. Translations, encoding space-time dependence, naturally emerge
as particular transformations inside the huge internal symmetry group. The possibility that large
matrices could dynamically generate the space-time has been also vigorously advocated in
\cite{bfss,ikkt}. In its original formulation the plaquette action for Wilson lattice theory simplifies to
\begin{equation}
S_{EK}(U_\mu)=-N\beta
\sum_{\mu\neq\nu}
{\rm Tr}\left (U_\mu U_\nu U_\mu^\dagger U_\nu^\dagger\right),
\label{ek}
\end{equation}
where $\mu,\nu=1,..,D$ and $U_\mu$ being $U(N)$ matrices. If the $U(1)^D$ symmetry of the
action (\ref{ek}) is not spontaneously broken, Schwinger-Dyson equations are unaltered under reduction process:
unluckily this is not always true in $D>2$, the equivalence holding only in the strong-coupling regime
\cite{neu}. The two-dimensional case has been studied in the eighties \cite{rossi} and  more recently
in \cite{nip}, leading to  contradictory conclusions.
\noindent
In order to avoid the problem with the limited viability of the model defined in eq. (\ref{ek}) it was
proposed, in even dimensions, the TEK model \cite{gonzo}
\begin{equation}
S_{TEK}(U_\mu)=-N\beta
\sum_{\mu\neq\nu}{\cal Z}_{\mu\nu}
{\rm Tr}\left (U_\mu U_\nu U_\mu^\dagger U_\nu^\dagger\right),
\label{tek}
\end{equation}
where the factor ${\cal Z}_{\mu\nu}$ is the ${\it twist}$,
\begin{equation}
{\cal Z}_{\mu\nu}={\cal Z}^*_{\nu\mu}=\exp(2\pi i k_{\mu\nu}/N).
\label{twist}
\end{equation}
Here $k_{\mu\nu}$ is an integer-valued antisymmetric matrix and $U_{\mu}$ is usually restricted to $SU(N)$.
The $U(1)^D$ symmetry is now reduced to $(Z_N)^D$ symmetry: it can be shown (see for example \cite{gonzo,das}) that
 Schwinger-Dyson equations still hold in the weak-coupling regime due to the non-trivial vacuum structure.
Numerical studies confirmed the equivalence with Wilson lattice gauge theory but the initial hope for
a deeper understanding of confinement properties were disappointed and the interest for the model
faded away.
\noindent
The situation changed recently thanks to a new interpretation of the TEK model as a non-perturbative
description of noncommutative gauge theory given by the action (we consider here the simplest case,
namely $U(1)$)
\begin{equation}
S=\frac{1}{4 g^2} \int d x^D F_{\mu\nu}(x)\star F^{\mu\nu}(x),
\label{ncac}
\end{equation}
where
\begin{equation}
F_{\mu\nu}(x)=\partial_{\mu} A_\nu-\partial_\nu A_\mu-i(A_\mu\star A_\nu
-A_\nu\star A_\mu).
\end{equation}
The star-product is defined as
\begin{equation}
f(x)*g(x)=\exp\left(i\frac{\theta^{\mu\nu}}{2}\frac{\partial}{\partial x^\mu}
\frac{\partial}{\partial y^\nu} \right)\left. f(x)g(y) \right|_{y=x},
\end{equation}
the noncommutative parameter $\theta^{\mu\nu}$ being an antisymmetric matrix with dimensions of a
length square. Noncommutative gauge invariance is realized through star-unitary transformations
$U(x)\star U(x)^\dagger=U(x)^\dagger \star U(x)=1$
\begin{equation}
A^\prime_{\mu}(x)=U(x)\star A_{\mu}(x)\star U(x)^\dagger+iU(x)\star\partial_{\mu}U(x)^\dagger.
\end{equation}
In refs. \cite{sm} it was shown that the lattice version of the action eq. (\ref{ncac})
turns out to be equivalent to certain reduced twisted $U(N)$ models at finite $N$: the formulation has
been given on a periodic lattice and the noncommutative parameter $\theta^{\mu\nu}$ is forced to take
discrete values. The continuum limit of lattice noncommutative gauge theory coincides precisely with the
large-$N$ limit of the reduced (twisted) model. The $SU(N)$ symmetry of the TEK action
corresponds to the gauge invariance in the noncommutative gauge theories, realized through the
star-unitary transformations. The novelty, observed in refs. \cite{sm}, is that one has to take the large-$N$
limit in a very peculiar way to land on noncommutative ground, starting from specific finite-$N$ TEK
formulations. We are going to discuss in details the two-dimensional case, that is our main concern here.
As we have anticipated in the introduction, we start from the very beginning, trying to exploit as much as
possible the algebraic properties of the "fuzzy" torus. The final results of our construction are equivalent
to the ones presented in \cite{sm}, but we feel that our approach is in some sense complementary and more suitable
to perform our computations.

\subsection{Gauge theories on a noncommutative toroidal lattice: the unconstrained formulation}
We start by recalling some elementary definitions and by introducing the basic formalism that we will extend
in the noncommutative case: a {\sl commutative} toroidal square lattice is a set of vectors of
the form
\begin{equation}
\label{eqq1}
{\vec{x}}= a\left (\begin{array}{c}n_1\\n_2\end{array}\right)\equiv a \vec{n},
\end{equation}
endowed with the equivalence relation $ {\vec{x}}\sim{\vec{y}}$~  {\rm iff}~
${\vec{x}}-{\vec{y}}=N a \vec{l}.$ Here $\vec{n}$ and $\vec{l}$  are integer-valued
vectors, while the parameter $a$ is identified with the lattice
spacing. A basis for the functions defined on this lattice is given by
\begin{equation}
\label{eqq4}
u_{\vec{k}}(\vec{x})=\exp(i \vec{k}\cdot \vec{x}),
\end{equation}
where the vector $\vec{k}$ belongs to the dual lattice
\begin{equation}
\label{eqq5}
\vec{k}= \frac{2\pi \vec{m}}{N a},
\end{equation}
$\vec{m}$ being a vector of integer numbers. The period of the momentum lattice is $2\pi/a$,
namely
\begin{equation}
\label{eqq6}
u_{\vec{k}+2\pi /a}(\vec{x})=u_{\vec{k}}(\vec{x}).
\end{equation}
The completeness of this basis is expressed by the condition
\begin{equation}
\label{eqq7}\frac{1}{N^2} \sum_{\vec{k}}
u^*_{\vec{k}}(\vec{x})u_{\vec{k}}(\vec{y})=
\delta^{P}_{{\vec{x}},\vec{y}},
\end{equation}
while the orthogonality relation is given by
\begin{equation}
\label{eqq8}\frac{1}{N^2} \sum_{\vec{x}}
u^*_{\vec{k}^\prime}(\vec{x})u_{\vec{k}}(\vec{x})=
\delta^{P}_{{\vec{k}},\vec{k}^\prime}.
\end{equation}
Here the symbols $\delta^{P}_{\vec{r},\vec{s}}$  denotes
the periodic Kronecker delta: this function  is equal to one when
$\vec{r}$ and $\vec{s}$ differ by any integer multiple of the
period of the lattice that they span and zero otherwise.
Any function $f(\vec{x})$  defined on this lattice can be Fourier-expanded
on this basis
\begin{equation}
\label{eqq9}
f(\vec{x})=\frac{1}{N}
\sum_{\vec{k}} f_{\vec{k}}~ u_{\vec{k}}(\vec{x}),
\end{equation}
where
\begin{equation}
\label{eqq10} f_{\vec{k}}=\frac{1}{N}\sum_{\vec{x}} f(\vec{x})~
u_{\vec{k}}(\vec{x}).
\end{equation}
The algebra of functions over a toroidal bidimensional lattice is
completely defined by the following property of this basis:
\begin{equation}
\label{commalg}
u_{\vec{k}}(\vec{x})u_{\vec{k^\prime}}(\vec{x})
=u_{\vec{k}+\vec{k^\prime}}(\vec{x})=
u_{\vec{k^\prime}}(\vec{x}) u_{\vec{k}}(\vec{x}).
\end{equation}
In fact,  by means of eq. (\ref{commalg}), the Fourier coefficients of
the product, $h(x)$, of two functions $f(x)$ and $g(x)$ are
\begin{equation}
h_{\vec{k}}=\frac{1}{N}\sum_{\vec{p}+\vec{q}=\vec{k}} f_{\vec{p}}~
g_{\vec{q}},
\end{equation}
i.e. the convolution of the original coefficients.
\noindent
Each element in the present basis is also an eigenstate of the translation
operators $T_1$ and $T_2$,
\begin{equation}
\label{commtrans}
T_{1,2}^\dagger u_{\vec{k}}(\vec{x}) T_{1,2} =u_{\vec{k}}(\vec{x}+a\vec{\delta}_{1,2})=
e^{i\vec{k}\cdot\vec{\delta}_{1,2}}
u_{\vec{k}}(\vec{x}),
\end{equation}
where $\vec{\delta}_1\equiv a (1,0)$ and $\vec{\delta}_2 \equiv a (0,1)$.

\noindent
Although the definition eq. (\ref{eqq1}) is the most simple for a
commutative lattice, it is not the natural one in the noncommutative
langauge where algebraic relations play a more fundamental role.
In this spirit, one can characterize a commutative square lattice
through  the abstract algebra eq. (\ref{commalg}) in a fixed point
\begin{equation}
\label{commalg1}
u_{\vec{k}}u_{\vec{k}^\prime}=u_{\vec{k}+\vec{k}^\prime}=
u_{\vec{k}^\prime}u_{\vec{k}}
\end{equation}
and then  one can reconstruct the basis in the other sites by means of the
condition eq. (\ref{commtrans}) written in the form
\begin{equation}
\label{commtrans1}
T_1^\dagger u_{\vec{k}} T_1 =
e^{i\vec{k}\cdot\vec{\delta}_1}
u_{\vec{k}}\ \ \ \
{\rm and}
\ \ \ \
T_2^\dagger u_{\vec{k}} T_2 =
e^{i\vec{k}\cdot\vec{\delta}_2}
u_{\vec{k}}.
\end{equation}

\noindent
The usual geometrical representation eq. (\ref{eqq1})
will appear when we try to realize explicitly eq. (\ref{commalg1}) and
eq. (\ref{commtrans1}). We shall call  ${\it noncommutative}$ square
lattice any representation of the deformed algebra
\begin{equation}
\label{nclattice}
{\cal U}_{\vec{k}} {\cal U}_{\vec{k}^\prime}=
\exp\left (\pi i\Theta(k_1 k^\prime_2 -k_2 k_1^\prime)\right)
{\cal U}_{\vec{k}+\vec{k}^\prime}=
\exp\left (2\pi i\Theta(k_1 k^\prime_2 -k_2 k_1^\prime)\right)
{\cal U}_{\vec{k}^\prime} {\cal U}_{\vec{k}}
\end{equation}
and of the relations
\begin{equation}
\label{nccommtrans}
T_1^\dagger {\cal U}_{\vec{k}} T_1 =e^{i\vec{k}\cdot\vec{\delta}_1}
{\cal U}_{\vec{k}}\ \ \ \
{\rm and}
\ \ \ \
T_2^\dagger {\cal U}_{\vec{k}} T_2 =e^{i\vec{k}\cdot\vec{\delta}_2}
{\cal U}_{\vec{k}}.
\end{equation}
Obviously, this is a sensible definition only if it is not affected by the
periodicity of the momenta $\vec{k}\to \vec{k}+2\pi /a$. This imposes
that
\begin{equation}
\label{Theta}
\frac{2 \pi^2 \Theta}{N a^2}=r \ \ \ \ {(\rm an~ integer)}\ \
\Rightarrow \Theta=\frac{N r}{2\pi^2} a^2.
\end{equation}
In order to study the representations of the algebra  (\ref{nclattice}), it is useful
to rewrite everything in terms of adimensional quantities, namely introducing explicitly the
integer vectors $\vec{m}$ defined in eq. (\ref{eqq5}). Now the
algebra  eq. (\ref{nclattice}) and the relations eq. (\ref{nccommtrans}) reads
\begin{eqnarray}
\label{adimalgebra}
&&{\cal U}_{\vec{m}}~ {\cal U}_{\vec{m}^\prime}=
e^{\pi i\theta(m_1 m^\prime_2-m_2 m^\prime_1
)}{\cal U}_{\vec{m}+\vec{m}^\prime}
=e^{2\pi i\theta(m_1 m^\prime_2-m_2 m^\prime_1)}
{\cal U}_{\vec{m}^\prime}~ {\cal U}_{\vec{m}},\nonumber\\
&&T_1^\dagger {\cal U}_{\vec{m}} T_1 =e^{2\pi i m_1/N}
{\cal U}_{\vec{m}}\ \ \ \
{\rm and}
\ \ \ \
T_2^\dagger {\cal U}_{\vec{m}} T_2 =e^{2\pi i m_2/N}
{\cal U}_{\vec{m}},
\end{eqnarray}
with
\begin{equation}
\label{theta}
\theta=2 r/N.
\end{equation}
\noindent
The representations of eqs. (\ref{adimalgebra}) can be built starting from the two generators
\begin{equation}
\label{U1U2}
\mathbb{U}_1={\cal U}_{(1,0)}\ \ \ \  \ \
{\rm and}\ \  \ \ \ \
\mathbb{U}_2={\cal U}_{(0,1)},
\end{equation}
which obey the Weyl-'t Hooft algebra  \cite{hooft}
\begin{equation}
\label{eaters}
\mathbb{U}_1\mathbb{U}_2=\exp(2\pi i 2r/N)\mathbb{U}_2\mathbb{U}_1
\end{equation}
as well as the constraints
\begin{equation}
\label{trasl1}
T_1^\dagger \mathbb{U}_1 T_1 =e^{2 \pi i/N} \mathbb{U}_1\ \ \ \
T_1^\dagger \mathbb{U}_2 T_1 = \mathbb{U}_2\ \ \ \
T_2^\dagger \mathbb{U}_1 T_2 = \mathbb{U}_1\ \ \ \
T_2^\dagger \mathbb{U}_2 T_2 =e^{2 \pi i/N} \mathbb{U}_2.
\end{equation}
The generic operator ${\cal U}_{\vec{m}}$ is then easily realized as
\begin{equation}
\label{Um}
 {\cal U}_{\vec{m}}= \exp( -2\pi i r~ m_1 m_2/N)
\mathbb{U}_1^{m_1} \mathbb{U}_2^{m_2}.
\end{equation}
We stress that we have reduced our original task, the construction of the noncommutative fuzzy torus, to
a well-defined algebraic problem, namely to find the representation of the algebra eq. (\ref{eaters}) and of
the relations eqs. (\ref{trasl1}).

\noindent
We first analyze the case of ${\it irreducible}$ representations. It is well-known there exists only one
irreducible representation of the algebra eq. (\ref{eaters}): denoting with $l={\rm gcd}\,(N,2r)$,
that can be characterized as follows ($2r'=2r/l$)
\begin{equation}
\mathbb{U}_1=(U_1^0)^{2r'}\ \ \ \ {\rm and} \ \ \ \  \mathbb{U}_2=U_2^0,
\label{eaters1}
\end{equation}
where $U_1^0$ and $U_2^0$ are the fundamental twist-eaters, satisfying
the basic relation ($N'=N/l$) $U_1^0 U_2^0 = \exp(2\pi i/N') U_2^0 U_1^0$. Its dimension is exactly $N'$.
In the following, to avoid useless complications, we limit ourselves to the case where $2r$ and $N$ are
coprime, namely $l=1$, $r'=r$, $N=N'$. Let us notice this means that our $N$ is odd.

\noindent
At this point we can also realize the translation operators $T_i$ in terms of the
fundamental twist-eaters. Let us define
\begin{equation}
\label{sol1}
T_1=(U_2^0)^s \ \ \ \ {\rm and}\ \ \ \ T_2=(U_1^0)^\dagger,
\end{equation}
where the integers $s$ and $k$ satisfies the Diophantine equation\footnote{
A solution to this equation  always exists since $ 2 r$ and $N$ are coprime.}
\begin{equation}
\label{sol1a}
(2 r) s- k N=1.
\end{equation}
Then $T_1$ obviously commutes with $\mathbb{U}_2$ and
\begin{equation}
T_1^\dagger \mathbb{U}_1 T_1=
(U_2^0)^{s\dagger} (U_1^0)^{2 r} (U_2^0)^s
=
e^{2 \pi i(2 r s )/N} \mathbb{U}_1=e^{2 \pi i(k N+1)/N} \mathbb{U}_1
=e^{2 \pi i/N} \mathbb{U}_1.
\end{equation}
In the same way it is manifest that $T_2$ commutes with $\mathbb{U}_1$,
while
\begin{equation}
T_2^\dagger \mathbb{U}_2 T_2=
U_1^0 U_2^0 (U_1^0)^\dagger
=
e^{2 \pi i/N} \mathbb{U}_2.
\end{equation}
It is intriguing to notice that the operators $T_i$ define a noncommutative
lattice whose parameter $\theta^\prime$ is
\begin{equation}
T_1 T_2=(U_2^0)^s (U_1^0)^\dagger=\exp(2\pi i  s/N) (U_1^0)^\dagger
(U_2^0)^s=\exp\left( 2\pi i\left(\frac{1+k N}{2 r N}\right)\right)
T_2 T_1.
\label{curva}
\end{equation}
The role of translation operators  is instead played by $\mathbb{U}_1^\dagger$
and $\mathbb{U}_2^\dagger$ respectively. These two tori can be mapped one into
the other through a particular transformation: let us define the matrix
\begin{equation}
\left(\matrix{a & b\cr
c & d}\right)=
\left(\matrix{-k N +4 r^2 &N (s-2 r) \cr k(2 r -s) &s^2-k N}\right)
\end{equation}
that belongs to $SL(2,\mathbb{Z})$, namely
\begin{equation}
a d-b c=(4 r^2 -k N)(s^2-k N)+N k(s-2 r)^2=(k N-2 r s)^2=1.
\end{equation}
This simple modular transformation maps the noncommutative parameter $\theta$ into
\begin{equation}
\theta^\prime=\frac{c+d\theta}{a+b\theta}=\frac{c N+ 2 r d}{a N +2 r b}=
\frac{s}{N}=\frac{1+k N}{2 r N},
\end{equation}
while the volume of the two lattices is trivially the same.
This is an example of the so-called Morita equivalence \cite{schwarz},
which has played a relevant role in the recent approaches to NCYM$_2$
\cite{gura2,gsv,pasz,pasz2}. How this symmetry emerges in the present
framework is discussed in appendix A.

\noindent
The translations constructed above are the only ones possible if we
strictly consider irreducible representations: the phase appearing in the commutation relation
eq. (\ref{curva}) can be naturally associated to the (necessarily constant) curvature
of the translation operators. Next we would like to construct
translations with an arbitrary constant curvature, namely such that
\begin{equation}
T_1 T_2=\exp(2\pi i  (m/n+s/N)) T_2 T_1=\exp(2\pi i (N m+s n)/(n N)) T_2 T_1,
\label{traslo}
\end{equation}
where $m$ and $n$ are arbitrary integers, while $s$ has been defined in eq. (\ref{sol1a}).
This is naturally accomplished if we accept reducible representations of the twist-eaters algebra:
in the general case the following relations have to be satisfied
\begin{eqnarray}
\label{pippo1}
\mathbb{U}_{1}\mathbb{U}_{2}&=&e^{2 \pi i (2 r)/N} \mathbb{U}_2\mathbb{U}_1,\nonumber\\
T_1 T_2&=&e^{2\pi i  (N m+s n)/(n N)} T_2 T_1,
\end{eqnarray}
and
\begin{eqnarray}
\label{pippo2}
T_1^\dagger \mathbb{U}_1 T_1 &=&e^{2 \pi i/N} \mathbb{U}_1,\ \ \ \
T_1^\dagger \mathbb{U}_2 T_1 = \mathbb{U}_2,\nonumber\\
T_2^\dagger \mathbb{U}_2 T_2 &=&e^{2 \pi i/N} \mathbb{U}_2,\ \ \ \
T_2^\dagger \mathbb{U}_1 T_2 = \mathbb{U}_1,
\end{eqnarray}
which can be seen as a four dimensional noncommutative torus, whose $\Theta$-matrix
is not in a canonical form. To represent the above structure, we find useful to introduce
the new translation operators $\tilde T_i$
\begin{equation}
\tilde T_1=\mathbb{U}_2^{\dagger s} ~ T_1,
\ \ \ \ {\rm and} \ \ \ \
\tilde T_2=\mathbb{U}_1 T_2.
\end{equation}
These operators satisfy a simpler algebra, namely
\begin{eqnarray}
\tilde T_1^\dagger \mathbb{U}_1 \tilde T_1&=&
T_1^\dagger ~\mathbb{U}_2^{ s} \mathbb{U}_1 \mathbb{U}_2^{\dagger s} ~ T_1
= e^{-2\pi i (2 r s)/N}T_1^\dagger  \mathbb{U}_1 T_1=
\mathbb{U}_1,\nonumber\\
\tilde T_2^\dagger \mathbb{U}_2 \tilde T_2&=&
T_2^\dagger ~\mathbb{U}_1^{\dagger} \mathbb{U}_2 \mathbb{U}_1 ~ T_2
= e^{-2\pi i /N}T_2^\dagger  \mathbb{U}_2 T_2
= \mathbb{U}_2,
\end{eqnarray}
where the Diophantine equation has been used.
Obviously we also have that
\begin{equation}
\tilde T_1^\dagger \mathbb{U}_2 \tilde T_1=\mathbb{U}_2
\ \ \ \
{\rm and}
\ \ \ \
\tilde T_2^\dagger \mathbb{U}_1 \tilde T_2=\mathbb{U}_1.
\end{equation}
All these equations can be summarized in the statement that the
operators $\tilde T_i$ commute with the coordinates $\mathbb{U}_i$.
The operators $\tilde T_i$ define an independent bidimensional torus, whose
noncommutative parameter $\theta$ is determined by
\begin{equation}
\tilde T_1 \tilde T_2=
e^{2\pi i((m N+s n)/(n N) -s/N)}\tilde T_2 \tilde T_1=
e^{2\pi i m/n}\tilde T_2 \tilde T_1.
\end{equation}
We finally end up with the announced reducible case: the description of the general
relations eq. (\ref{pippo1}) and eq. (\ref{pippo2}) is encoded into two disjoint
noncommutative tori, whose $\theta$'s are respectively
\begin{equation}
\theta=2 r/N \ \ \ \ {\rm and} \ \ \ \ \theta_T=m/n.
\end{equation}
The cheapest way to obtain a representation of the complete algebra is, in fact,
to take the tensor product of the representations of the two tori. The dimension
of a generic representation is a multiple of $(n^\prime N)$ where
$n^\prime=n/{\rm gcd}(n,m)$.

\noindent
A better understanding of the geometrical meaning of this construction
can be obtained using the following  parametrization for the integers
$m$ and $n$,
\begin{equation}
n=N p- (2 r) q \ \ \ \ \ \  m=-s p+k q,
\label{msun}
\end{equation}
where $p$ and $q$ are arbitrary integers. This parametrization is absolutely
general being inverted by
\begin{equation}
p= -2 r m-k n
\ \ \ \ \ \
q=-m N- n s.
\end{equation}
These two equations also imply the interesting property that the
$\textrm{gcd}(m,n)=\textrm{gcd}(p,q)$.  The geometrical data ${\cal N}=\textrm{gcd}(m,n)=\textrm{gcd}(p,q)$
can be identified  with the rank of the gauge group of the noncommutative theory \cite{pasz}.

\noindent
In our approach $n$ is a multiple of the dimension of the representation
of the second
torus, therefore it is a positive number; in terms of $p$ and $q$ this
condition
reads
\begin{equation}
n=N p -(2r) q=N(p-\theta q)>0.
\label{posi}
\end{equation}
This condition, in the continuum, will appear as the positive cone constraint and will
determine $(p,q)$ associated to inequivalent projective modules.
The value of the $\theta$ of the original $T_i$ operators is instead
\begin{equation}
\frac{N m+ s n}{n N}=-\frac{q}{N^2(p-\theta q)}.
\end{equation}
We have now to exhibit an explicit representation of eq. (\ref{pippo1}): the natural choice is
given by
\begin{equation}
\mathbb{U}_1= (U_1^0)^{2r}\otimes \mathbb{I}_{n^\prime},\ \ \ \
\mathbb{U}_2= (U_2^0)\otimes \mathbb{I}_{n^\prime},
\end{equation}
\begin{equation}
T_1= (U_2^0)^{s}\otimes \Gamma_1^{m^\prime},\ \ \ \
T_2= (U_1^0)^\dagger\otimes \Gamma_2,
\end{equation}
where $\Gamma_1\Gamma_2=\exp(2\pi i /n^\prime) \Gamma_2\Gamma_1$ and
$m^\prime=m/{\rm gcd}(n,m)$.

\noindent
In the following we shall interpret the two integers $p$ and $q$ introduced
above as the same $p$ and $q$ that classify the modules on the noncommutative
torus and we shall construct the gauge theories on these modules. Let us
start with the simple trivial module, namely that with vanishing $q$
and $p=1$:

\noindent
{\bf (1,0)}: We choose $ n=n'= N$ and $m=m'=-s$. In this module the fundamental
translation operators commutes and their form is simply
\begin{equation}
T_1= (U_2^0)^{s}\otimes (U_2^0)^{-s}\ \ \ \
T_2= (U_1^0)^\dagger\otimes (U_1^0)^\dagger.
\end{equation}
Then the  coordinates  $\mathbb{U}_i$ are
\begin{equation}
\mathbb{U}_1= (U_1^0)^{2r}\otimes \mathbb{I}_N,\ \ \ \
\mathbb{U}_2= (U_2^0)\otimes \mathbb{I}_N.
\end{equation}
To define a gauge theory we must construct the most general
translation operators, which satisfy the algebra eq. (\ref{pippo1})
and eq. (\ref{pippo2}) except for the constant curvature condition.
We call these operators $D_i$ and they are simply given by
\begin{equation}
D_1=(U_2^0)^{s}\otimes V_1\ \ \ \
D_2= (U_1^0)^\dagger\otimes V_2,
\end{equation}
where $V_1$ and $V_2$ are arbitrary $N\times N$ unitary matrices, encoding the degrees of
freedom of the related gauge theory.

\noindent
The next step, in our construction, is to write down an action: the natural one is the
generalized Wilson one-plaquette action, whose form is
\begin{equation}
S=\alpha-\beta \Bigl(\exp(2 i\phi){\rm Tr}[D_1 D_2 D_1^\dagger D_2^\dagger]+
 \exp(-2i\phi){\rm Tr}[D_2 D_1 D_2^\dagger D_1^\dagger]\Bigr).
\end{equation}
The parameters $\alpha$ and $\beta$ are real numbers, while $\phi$ is a background
phase\footnote{The introduction of a background phased is suggested by the possible presence
 of a background flux in the continuum limit}. The constant $\alpha$ is fixed by requiring that $S$ be positive
definite. The action can be written in the following equivalent
manner
\begin{equation}
S=\beta {\rm Tr}\Bigl[\Bigl(\exp(i\phi) D_1 D_2-\exp(-i\phi)  D_2  D_1\Bigr)
\Bigl(\exp(-i\phi) D_2^\dagger D_1^\dagger-\exp(i\phi)  D_1^\dagger  D_2^\dagger\Bigr)\Bigr]
+\alpha-2 \beta N^2,
\end{equation}
suggesting the simple choice
\begin{equation}
\alpha=2 \beta N^2,
\label{sub}
\end{equation}
combined with the requirement that $\beta>0$. We arrive therefore to the nice expression
\begin{equation}
S=\beta {\rm Tr}\Bigl[\Bigl(\exp(i\phi) D_1 D_2-\exp(-i\phi)  D_2  D_1\Bigr)
\Bigl(\exp(-i\phi) D_2^\dagger D_1^\dagger-\exp(i\phi)  D_1^\dagger  D_2^\dagger\Bigr)\Bigr],
\end{equation}
where $\phi$ is still undetermined: we choose the background phase by requiring that the absolute minimum,
i.e. the one with zero action, corresponds to the derivatives with constant curvature on the module, in
complete analogy with the continuum description. In our case, due to the fact we have commuting derivatives,
$\phi=0$ modulo $N$. The final action turns out to be
\begin{eqnarray}
\label{plaquette12}
&&\!\!\!\!\!\!
S=\beta {\rm Tr}\Bigl[\Bigl(D_1 D_2-D_2  D_1\Bigr)\Bigl(D_2^\dagger D_1^\dagger- D_1^\dagger D_2^\dagger\Bigr)\Bigr]=\\
&&\!\!\!\!\!\! =\beta N{\rm Tr}\Bigl[\!\Bigl(\exp(\pi i s/N)V_1
V_2\!-\exp(-\pi i s/N)V_2 V_1\Bigr)\!\! \Bigl(\exp(-\pi i
s/N)V_2^\dagger V_1^\dagger\!-\exp(\pi i s/N)V_1^\dagger
V_2^\dagger\Bigr)\!\Bigr].\nonumber
\end{eqnarray}
In terms of the $V_i$ matrices the theory is the well-known TEK model, described before,
with a twist factor given by $\exp(2\pi i s/N)$. It is important to notice the appearing
of the factor $N$ in front to the classical action: it carries part of the space-time dependence,
being in fact factored out as the volume of the translations.

\noindent
We are ready now to consider the general case.

\noindent
{\bf (p,q)}: For  simplicity, we  shall take  $p$ and $q$ to be coprime
or equivalently the gauge group to be $U(1)$, then
$n=N p-2 r q$ and $m=k q -p s$ are coprime. In this case the
fundamental  translation operators are represented as
\begin{equation}
T_1= (U_2^0)^{s}\otimes \Gamma_1^{k q -p s}\ \ \ \
T_2= (U_1^0)^\dagger\otimes \Gamma_2,
\end{equation}
where the dimension of $\Gamma_i$ is $n=N p-2 r q$ since
gcd$(m,n)={\rm gcd}(p,q)=1$.
The most general derivatives on this module can be written as
\begin{equation}
D_1=(U_2^0)^{s}\otimes V_1\ \ \ \
D_2= (U_1^0)^\dagger\otimes V_2,
\end{equation}
where now the matrices $V_i$ are $(N p-(2 r q))\times  (N p -(2 r q))$ unitary
matrices. The corresponding action  is again
\begin{equation}
S=\beta {\rm Tr}\Bigl[\Bigl(\exp(i\phi) D_1 D_2-\exp(-i\phi)  D_2  D_1\Bigl)
\Bigr(\exp(-i\phi) D_2^\dagger D_1^\dagger-\exp(i\phi)  D_1^\dagger  D_2^\dagger\Bigr)\Bigr],
\label{plaquette13}
\end{equation}
where $\phi$ is now different from zero modulo $N$ and  determined by the
condition that the absolute minimum is the derivation corresponding to the
constant curvature derivative. Its value is
\begin{equation}
\phi=-\pi (m/n+s/N)\ \ \ \ {\rm modulo}\  N.
\label{connec}
\end{equation}
In complete analogy with the trivial case, we can perform part of the traces,
writing the action in terms of the fundamental variables $V_1,V_2$ carrying the dynamical degrees
of freedom. In doing  so $\beta$ is again renormalized by a factor $N$, a crucial feature when discussing
the continuum limit. We end up with a TEK action twisted by the phase
\begin{equation}
\phi'=-\pi\frac{m}{n}=-\pi\frac{k q-sp}{Np-2 r q},
\end{equation}
while the dimension of the matrices is equal to $Np-2 r q$, {\it i.e.}
\begin{equation}
\label{plaquette15}
S=\beta N{\rm Tr}\Bigl[\Bigl(\exp( i \phi^\prime)V_1 V_2-\exp(- i  \phi^\prime)V_2 V_1\Bigr)
\Bigl(\exp(- i \phi^\prime)V_2^\dagger V_1^\dagger-\exp( i  \phi^\prime)V_1^\dagger
V_2^\dagger\Bigr)\Bigr].
\end{equation}
This is, of course, the final result we were looking for
and it deserves some comments. First of all we remark that we were able to write the discretized theory
in terms of a "conventional" TEK model, in the sense no constraint is required on the matrices $V_i$. Once we have
fixed the number of lattice points, $N^2$, and the noncommutative (adimensional) parameter, $\theta=2r/N$, all
the $(p,q)$ modules satisfying the positivity constraint eq. (\ref{posi}) are simply described. The topological
distinction is effectively encoded on the size of the matrices and on the twist-phase characterizing the
matrix model action. The second point we would like to stress is concerning our actual choice of the background
phase: we have explicitly tuned the absolute minimum of the classical action with the constant curvature connection
of the associated (discretized) projective module, a well-known property of the continuum description \cite{cr}.
We alert the reader that other possibilities are in principle allowed, as we will see, consistently with recovering
this crucial property in the continuum limit. Finally we stress the explicit appearance of the factor $N$
in front of the action, due to a partial decoupling of the space-time structure in taking the traces: it will play
a relevant role in constructing the continuum limit.

\subsection{The equivalence with the constrained formulation}
In the literature, another TEK model describing the Yang-Mills theory in two dimensions over the generic
module $(p,q)$, at discretized level, has been discussed in ref. \cite{sm}. The model proposed there differs,
apparently, in a fundamental aspect from the one we derived here: it is a ${\it constrained}$ TEK model, ${\it i.e.}$
the matrices entering in the classical action must satisfy a non-trivial equation. Although the two models
are required to be relatively consistent only in the continuum limit, it is nevertheless important to find their
precise relation directly at discretized level.

\noindent
In order to facilitate the comparison between our action and theirs, we shall begin by translating their model in our
notation. Their starting point is a noncommutative lattice of dimension $N= \tilde n q$,
generated by the matrices satisfying the Weyl-'t Hooft algebra
\begin{equation}
\label{C3}
U_1 U_2=e^{2\pi i \theta} U_2 U_1,
\end{equation}
with\footnote{In the following, we shall assume gcd$(p \tilde n-\tilde m, \tilde n q)=1$}
\begin{equation}
\theta =\frac{p}{q}-\frac{\tilde m}{\tilde n q}=\frac{p \tilde n- \tilde m}{\tilde n q}.
\end{equation}
The matrix action, defining their final model, is written as
\begin{equation}
\label{tekcons}
S_1=-\beta Z \mathrm{Tr}(D_1 D_2 D_1^\dagger D_2^\dagger)-\beta Z^*\mathrm{Tr}(D_2 D_1 D_2^\dagger D_1^\dagger),
\end{equation}
where the matrices $D_i$ are subjected to the constraints
\begin{eqnarray}
\label{cons}
D_1^\dagger U_1 D_1 &=&e^{2\pi i /(\tilde n q)} U_1, \ \ \
D_2^\dagger U_1 D_2 =  U_1,\nonumber\\
D_2^\dagger U_2 D_2 &=&e^{2\pi i /(\tilde n q)} U_2,\ \ \
D_1^\dagger U_2 D_1 = U_2.
\end{eqnarray}
The matrices $D_i$ are therefore generators of translations. The parameter $\beta$ is an overall normalization,
while $Z$ is a phase chosen, at the end of the day, to select the desired vacuum: we see the actions
eq. (\ref{plaquette13}) and eq.  (\ref{tekcons}) are formally the same (a suitable subtraction can be
easily performed in eq. (\ref{tekcons}) to make the action positive defined).
The difference arises when it comes to representing
explicitly the algebra eq. (\ref{C3}) and the matrices $D_i$ satisfying the constraints eq. (\ref{cons}).

\noindent
The algebra eq. (\ref{C3}) has been realized, in ref. \cite{sm}, by means of matrices
$(\tilde m \tilde n q^2)\times (\tilde m \tilde n q^2)$ given by
\begin{equation}
\label{repszabo}
U_1=(\Gamma_2)^{\tilde m}\otimes (\tilde\Gamma_1^\dagger)^p,
\ \ \ \ U_2=(\Gamma_1)^{\tilde m}\otimes(\tilde\Gamma_2^\dagger)
\end{equation}
with
\begin{equation}
\label{cirio1}
\Gamma_1 \Gamma_2 =e^{2\pi i/(\tilde m \tilde n q)} \Gamma_2 \Gamma_1\ \ \ \ \
 \tilde\Gamma_1\tilde \Gamma_2 =e^{2\pi i/q} \tilde\Gamma_2\tilde\Gamma_1,
\end{equation}
and $(U_i)^{\tilde n q}=\mathbb{I}$.
The matrices $\Gamma_i$ and $\tilde \Gamma_i$ are the irreducible representations of the algebras eq. (\ref{cirio1})
and thus of dimensions $(\tilde m \tilde n q)\times (\tilde m \tilde n q)$ and $q\times q$ respectively.
The solutions of eq. (\ref{cons}) are then parameterized in terms of a particular $\bar D_i$,
\begin{equation}
\bar D_1=(\Gamma_1)^\dagger\otimes \mathbb{I}_q\ \ \ \ \  \bar D_2=\Gamma_2\otimes \mathbb{I}_q,
\end{equation}
solving the equations, and the matrices that commute with $U_i$.
In other words, the translation operators $D_i$ are written as
\begin{equation}
\label{cirio3}
D_i =S_i \bar D_i\ \ \ \ \ \mathrm{where} \ \ \ \  S_i U_j=U_j S_i \ \ \ \ \forall\ \  i,j=1,2.
\end{equation}
The set of the matrices commuting with the generators $U_i$  of the lattice are then parameterized with the help
of two other generators $Z_i$,
\begin{equation}
Z_1=(\Gamma_2)^{\tilde n}\otimes (\tilde\Gamma_1)^\dagger\  \ \ \ \ \ \ \ Z_2=(\Gamma_1)^{\dagger \tilde n}
\otimes (\tilde\Gamma_2)^{\hat a},
\end{equation}
with $a$ defined by the Diophantine equation
$
\hat a p+b q=1.
$
It is not difficult to verify that $Z_i U_j=U_j Z_i$ and
\begin{equation}
\label{cirio2}
Z_1 Z_2=e^{2\pi i \theta^\prime } Z_2 Z_1,\  \ \ \ \ \mathrm{with}\ \ \ \ \
\theta^\prime= \frac{\hat a\theta+b}{p-q\theta}.
\end{equation}
We have also $(Z_i)^{\tilde m q}=\mathbb{I}$.
The most general $S_i$  is therefore a unitary  matrix  belonging to the algebra generated by eq. (\ref{cirio2}).
Finally, one completes the definition of model by formally expressing  the operators $D_i$  in terms of
unconstrained variables $Z_i$ through the relation eq. (\ref{cirio3}). The procedure, however, cannot be
carried out explicitly without loosing the elegant and simple structure of the matrix model. This feature
makes not straightforward to use this model for concrete computations on the torus. Nevertheless, this
TEK representation admits an elegant translation as the Wilson action living on the dual noncommutative
lattices of the derivation endowed with the (discretized) star-product defined by $\theta^\prime$.

\noindent
Our procedure would have, instead, produced a smaller unconstrained TEK model,
with matrices $V_i$ of dimension $n\times n$,
\begin{equation}
n=N p-2 r q= \tilde n q p - (p\tilde  n -\tilde m) q= \tilde m q,
\end{equation}
and a twist-phase
\begin{equation}
\phi=\frac{m}{n}=-\frac{s}{\tilde n q}-\frac{1}{\tilde m \tilde n q}.
\end{equation}
We have used here our Diophantine equation, that in terms of the new variables is
\begin{equation}
(p \tilde n-\tilde m) s- k \tilde n q=\tilde n(p s-k q)-\tilde m s=-\tilde n m-\tilde m s=1.
\end{equation}
Then its action is
\begin{equation}
\label{S2}
S_2=\beta \tilde n q{\rm Tr}\Bigl[\Bigl(\exp(\pi i\phi )V_1 V_2-\exp(-\pi i \phi)V_2 V_1\Bigr)
\Bigl(\exp(-\pi i \phi)V_2^\dagger V_1^\dagger-\exp(\pi i \phi )V_1^\dagger
V_2^\dagger\Bigr)\Bigr].
\end{equation}
We shall try to answer the question on how these two models are related:
in particular we shall explain that there is a change of variables that reduces the action
eq. (\ref{tekcons}) to the action eq. (\ref{S2}).
\noindent
The key point to notice is the following: we know, from the general theory, that there exists only one
irreducible representation of the Weyl-'t Hooft algebra eq. (\ref{C3}) and that its dimension is $\tilde n q$.
The unitary representation eq. (\ref{repszabo}) is then completely reducible and it can be written in the form
\begin{eqnarray}
U_1&=&(\Gamma_2)^{\tilde m}\otimes (\tilde\Gamma_1^\dagger)^p
=\Omega^\dagger(\bar U_1^{p \tilde n-\tilde m} \otimes \mathbf{d}^1_{\tilde mq}) \Omega\nonumber\\
U_2&=&(\Gamma_1)^{\tilde m}\otimes(\tilde\Gamma_2^\dagger)= \Omega^\dagger(\bar U_2 \otimes
\mathbf{d}^2_{\tilde mq})\Omega,
\end{eqnarray}
where $\bar U_i$ are the irreducible representation of the algebra $\bar U_1 \bar U_2=e^{2\pi i/(\tilde n q)}
\bar U_2 \bar U_1$. Moreover $\mathbf{d}^i_{\tilde m q}$ are two unitary diagonal matrices of dimension
$\tilde m q\times \tilde m q$ and $\Omega$ is a unitary transformations: we shall show that each eigenvalue of the matrices
$U_i$ is degenerate $\tilde m q$ and therefore the matrices $\mathbf{d}^i_{\tilde m q}$ are actually proportional
to the identity $\mathbb{I}_{\tilde m q}$.

\noindent
To begin with, we consider  the generator $U_1$ and the matrix $Z_1$ commuting with it.  Let $e_1$ be one
of the common eigenvectors,
$$
U_1 e_1=e^{2\pi i \alpha} e_1 \ \ \ \ \mathrm{and}\ \ \ \
Z_1 e_1=e^{2\pi i \beta } e_1,
$$
then the vectors $f_i= Z_2^{i-1} e_1$ ($i=1,\dots,\tilde m q$) are $\tilde m q$ independent vectors. In fact,
the algebra eq. (\ref{cirio2}) implies that $f_i$ are eigenvectors of $Z_1$ corresponding to different eigenvalues.
On the other hand, since $Z_2$ commutes with $U_1$, the vectors $f_i$ are also eigenvectors of $U_1$ all with
the same eigenvalue $e^{2\pi i\alpha}$. Thus each eigenvalue of $U_1$ is degenerate at least $\tilde m q$. To
conclude that each eigenvalue is exactly $\tilde m q$ degenerate, we recall that the matrix $U_1$ has dimension
$(\tilde m \tilde n q^2)\times (\tilde m \tilde n q^2)$ and that  it possesses at least $\tilde n q $  different
eigenvalues because of the algebra eq. (\ref{C3})\footnote{That two matrices $U_i$ satisfying  the Weyl-'t Hooft algebra,
$$
U_1 U_2=e^{2\pi i m/n} U_2 U_1 \ \ \ \ \mathrm{with}\ \ \ \ \mathrm{gcd}(m,n)=1,
$$
possess at least $n$ distinct eigenvalues, is a trivial consequence of the following observation. Given an
eigenvector $e_1$ of $U_1$, the vectors $(U_2)^{i-1} e_1$ ($i=1,\dots,n$) are still eigenvectors of $U_1$ with
different eigenvalues because of the Weyl-'t Hooft algebra. In particular, in the only irreducible representation,
which has dimension $n$, all the eigenvalues are non-degenerate.}. A similar reasoning allows us to reach the same
result for $U_2$, obtaining therefore
\begin{eqnarray}
U_1&=& (\Gamma_2)^{\tilde m}\otimes (\tilde\Gamma_1^\dagger)^p
=e^{i\phi_1}\Omega^\dagger(\bar U_1^{p\tilde n-\tilde m} \otimes \mathbb{I}_{\tilde mq}) \Omega,\nonumber\\
U_2&=&(\Gamma_1)^{\tilde m}\otimes(\tilde\Gamma_2^\dagger)=  e^{i\phi_2}\Omega^\dagger(\bar U_2 \otimes
\mathbb{I}_{\tilde mq})\Omega.
\end{eqnarray}
This, in turn, implies that the matrices $S_i$ must be of the form
\begin{equation}
S_i= \Omega^\dagger(\mathbb{I}_{\tilde n q}\otimes \mathcal{V}_i) \Omega,
\end{equation}
where $\mathcal{V}_i$ are unconstrained unitary matrices of dimension $(\tilde m q)\times (\tilde m q)$.
The generic derivation can be then written
\begin{equation}
D_i=S_i \bar D_i=\Omega^\dagger(\mathbb{I}_{\tilde n q}\otimes \mathcal{V}_i) \Omega \bar D_i=
\Omega^\dagger(\mathbb{I}_{\tilde n q}\otimes \mathcal{V}_i) \bar\mathcal{D}_i \Omega,
\end{equation}
where $\bar\mathcal{D}_i=\Omega\bar D_i \Omega^\dagger$. The matrices $\bar\mathcal{D}_i$ are a
particular solution of the constraints eq. (\ref{cons}) with
\begin{equation}
\bar U_1^{p\tilde n-\tilde m} \otimes
\mathbb{I}_{\tilde mq}\equiv \mathcal{U}_1,\ \ \ \ \ \
\bar U_2\otimes
\mathbb{I}_{\tilde mq}\equiv \mathcal{U}_2.
\end{equation}
The background translations $\bar\mathcal{D}_i$ possess the same factorized structure of the
coordinates $\mathcal{U}_i$: this can be shown by writing $\bar\mathcal{D}_i$ as
follows
\begin{equation}
\bar\mathcal{D}_1=(\bar U_2^{s}\otimes \mathbb{I}_{\tilde m q})\mathrm{R_1}\ \ \ \ \ \
\bar\mathcal{D}_2=(\bar U_1^{\dagger}\otimes \mathbb{I}_{\tilde m q})\mathrm{R_2},
\end{equation}
where the first term is an alternative solution of the constraint (\ref{cons}) for the
$\mathcal{U}_i$. This, in turn, implies that the matrices $\mathrm{R_i}$ commutes with the
$\mathcal{U}_i$ and thus they can be written as
\begin{equation}
\mathrm{R_i}=\mathbb{I}_{\tilde n q}\otimes \mathcal{W}_i.
\end{equation}
In other words, we have shown that the matrices $\bar\mathcal{D}_i$ are of the form
\begin{equation}
\bar\mathcal{D}_1=\bar U_2^{s}\otimes \mathcal{W}_1\  \ \ \ \ \ \
\bar\mathcal{D}_2 =\bar U_1^{\dagger}\otimes \mathcal{W}_2,
\end{equation}
with $\mathcal{W}_1\mathcal{W}_2=e^{-2\pi i m/(\tilde m q)}\mathcal{W}_2\mathcal{W}_1$
since $\bar D_1 \bar D_2= e^{-2\pi i/(\tilde m\tilde n\tilde q)}
\bar D_2 \bar D_1$. The general derivation can be then parameterized as follows (we obviously have defined
$\mathcal{D}_i=\Omega D_i \Omega^\dagger$)
\begin{equation}
D_1=\Omega^\dagger\mathcal{D}_1\Omega=\Omega^\dagger(\bar U_2^{s}\otimes \mathcal{W}_1\mathcal{V}_1)\Omega\  \ \ \ \ \ \
D_2=\Omega^\dagger\mathcal{D}_2\Omega =\Omega^\dagger(\bar U_1^{\dagger}\otimes \mathcal{W}_2\mathcal{V}_2)\Omega,
\end{equation}
Substituting this representation in the action eq. (\ref{tekcons}), it reduces to
\begin{equation}
\label{tekcons1}
S_1=-\beta\, (\tilde n q )Z\ e^{2\pi i s/(\tilde n q)} \mathrm{Tr}
\Bigl[\mathcal{W}_1 \mathcal{V}_1 \mathcal{W}_2\mathcal{V}_2
\mathcal{V}_1^\dagger \mathcal{W}_1^\dagger\mathcal{V}_2^\dagger \mathcal{W}_2^\dagger\Bigr]+\mathrm{c.c.}.
\end{equation}
Changing variables to $V_i=\mathcal{V}_i\mathcal{W}_i$ and keeping in mind that $Z$ was
chosen in \cite{sm} as $Z=e^{-2\pi i/(\tilde m \tilde n q)}$, this action becomes identical
to that in eq. (\ref{S2}) up to the background phase and a (trivial) subtraction term. One can check
that the basic difference corresponds to a different choice of the configuration of minimal action.

\noindent
Summarizing, the models are completely equivalent. They describe the same physics in different basis.
In particular, the constrained basis of ref. \cite{sm} is the natural one for obtaining a (discretized)
star-product interpretation of the model while ours is the most suitable for performing the matrix model
computations, being in fact completely unconstrained.

\section{The continuum limits}
\noindent
The next step of our construction is, of course, to recover the continuum limit:
we will try to discuss the problem from a general point of view, having in mind the differences
with the ordinary case and hoping to elucidate some subtle points. We start by recalling
the basic relations between the size of our lattice, that we denote by $L$, and the dimensionful $\Theta$ parameter
\begin{equation}
\label{Theta2}
\Theta=\frac{N r}{2\pi^2} a^2=\frac{2r}{N}\frac{L^2}{4\pi^2},
\end{equation}
\begin{equation}
\label{size}
L=Na.
\end{equation}
Noncommutativity necessary implies that $\Theta$ has to be finite as $N$ becomes large: we see
that the situation drastically changes if we also require $L$ to be finite (noncommutative torus)
or not (noncommutative plane), reflecting into a non-trivial scaling of the parameter $r$. Finiteness
of $L$, in fact, determines the continuum limit ($a\to 0$) as
\begin{equation}
a\simeq \frac{L}{N}:
\label{equation}
\end{equation}
the finiteness of $\Theta$, therefore, implies that as $N\to\infty$
\begin{equation}
r\simeq(\frac{2\pi^2\Theta}{L^2})N,
\end{equation}
recovering the announced scaling of $r$. 
To reproduce the noncommutative plane we have, instead,
to send $L$ to infinity: this can be generically obtained by
scaling for large $N$
\begin{equation}
r\simeq N^{1-\gamma}\,\,\,\Rightarrow \,\,\, a\simeq\frac{1}{N^{1-\frac{\gamma}{2}}},
\end{equation}
$0\leq\gamma\leq 1$. At this level, all these limits are
potentially good in recovering the noncommutative plane. Usually
people considered $\gamma=1$ \cite{sm,nisj}: it corresponds
exactly  to the limits explored in \cite{luis,gura2,gsv}, where
the size of the noncommutative torus has been assumed to scale as
$\sqrt{N}$ in studying gauge theories. Actually, in the case of
scalar theories in two dimensions, the possibility of more general
scalings has been suggested in \cite{wmr}: there, in order to cope
with a solvable matrix model, the limit $\Theta\to\infty$ was also
taken in a correlated way. The authors pointed out the possible
appearance of three different phases, depending on the chosen
scaling: more recently the scalar case was examined in \cite{lsz}
by means of a powerful matrix model technique, in the
large-$\Theta$ limit too, and the existence of "exotic" scalings
was also discussed. Related investigations in Yang-Mills theory have been performed in 
\cite{kf}, where the possibility to consider more general scalings has been 
exploited in computing correlators of open Wilson lines. 
In the following we will restrict ourselves to
the canonical case $\gamma=1$, leaving the more general
possibility to future investigations. Within this choice the partition 
function and Wilson lines correlators appear to be dominated by the 
classical solutions of the system \cite{gsv}, consistently with localization 
theorems \cite{pasz}.

\noindent The unexpected result of these analysis is that, in
order to keep the noncommutative parameter finite as the lattice
spacing goes to zero, we need forcing $N\to\infty$ simultaneously
with $a\to 0$. At perturbative level this implies, generically,
that non-planar diagrams no longer vanish: the usual proof of
equivalence between TEK and Wilson lattice theory indeed requires
that only planar diagrams survive the large-$N$ limit and this is
achieved by first taking $N$ large and then going to continuum
limit. We are searching here, instead, for a {\it double scaling}
limit and, in particular, we are looking for non-perturbative
effects studying the behavior of the classical solutions of the
model.

\noindent  A first interesting consequence is the following: in
the non-compact case the space-time noncommutativity forces to
correlate the ultraviolet ($a\to 0$) limit with the infrared
($L\to\infty$) limit: we have therefore a non-perturbative
interpretation of the famous IR/UV connection, as pointed out in
\cite{sm}. In the compact case the double scaling procedure has
not a direct IR/UV interpretation: we have instead, in order to
stay with $\theta$ finite, to perform a limit on the $r$ parameter.
Quite surprisingly this offers the possibility to recover, as a
particular case, the commutative theory by simply tuning the limiting
value of $r$. We will return on this opportunity in Sect.5,
when discussing the classical solutions in the continuum limit.

\noindent Coming back to eq. (\ref{plaquette12}) we see that, of
course, the lattice spacing $a$ does not appear explicitly in the
definition of the matrix model: we have to identify the relation
between $\beta$ and $a$, establishing in this way the explicit
double scaling limit to be performed in the TEK model. Dimensional
considerations suggest a canonical scaling
\begin{equation}
\beta\sim \frac{1}{g^2a^2},
\label{scala}
\end{equation}
$g^2$ being the physical coupling constant: we remark that in
$D=2$ the coupling of the noncommutative Yang-Mills theory has the
dimension of a mass square. This choice deserves some comments.
We recall, first of all, that the canonical scaling correctly
reproduces the continuum limit of $2D$ ${\it commutative}$ lattice
gauge theories. TEK models are also seen to be equivalent to
gauge theories on particular periodic lattices, written in the Wilson form
through the discretized version of the noncommutative star-product \cite{sm}.
It is quite natural, therefore, to assume eq. (\ref{scala}) in defining the
continuum limit and we will adopt this point
of view in the rest of the paper. This choice clearly determines
the explicit form of the double scaling limit to be performed:
\begin{eqnarray}
&\beta\simeq N^2&\,\,\,\,\,\,\,\,\,\,{\rm NC\,\, torus},\nonumber\\
&\beta\simeq N&\,\,\,\,\,\,\,\,\,\,{\rm NC\,\,plane}.
\label{scali}
\end{eqnarray}
Having identified the precise form of the double scaling limit,
we can try to get some intuitions on its physical consequences:
we exploit the claimed equivalence of conventional TEK models with
commutative large-$N$ Yang-Mills theories. To this aim, let us review some well-known
facts about the two dimensions. \noindent Two-dimensional lattice gauge
theories, at large $N$, are known to have a non-trivial phase
structure \cite{gw}: a third-order phase transition occurs at
$\beta=1/2$, distinguishing a strong-coupling regime ($\beta<1/2$)
from the (physical) weak-coupling regime ($\beta>1/2$). The phase
transition is reflected by a different functional form of the
dimensionless string tension $k(\beta)$:
\begin{eqnarray}
k(\beta)&&=-{\rm ln}\,\beta\,\,\,\,\,\,\,\,\,\,\,\,\,\,\,\,\,\,\,\,\,\beta<1/2\nonumber\\
k(\beta)&&=-{\rm ln}\,(1-\frac{1}{\beta})\,\,\,\,\,\beta>1/2.
\label{fase}
\end{eqnarray}
The fact that the Wilson loops follows an exact area law, in the continuum limit, implies that one should tune
the bare coupling constant $\beta$, as the continuum limit is approached ($a\to 0$), using the second relation in
eq. (\ref{fase})
\begin{equation}
a^2=-\frac{1}{g^2}{\rm ln}\,(1-\frac{1}{\beta}),
\end{equation}
recovering the asymptotic behavior anticipated in eq.
(\ref{scala}). From the point of view of ordinary gauge theories,
the double scaling limit eq. (\ref{scali}) we have to consider
goes deeply into the weak-coupling phase: therefore it is not
expected to give new results in the Wilson theory, the extreme
weak-coupling phase being dominated by the trivial classical
solution $U=1$ (in axial gauge). On the other hand, the situation
should be rather different when our double scaling limit is
applied to the TEK model: its weak-coupling regime has to be somehow
different from conventional Wilson theory, at least at large
$\beta$, if noncommutative theories have to be reproduced. We
expect an highly non-trivial effect of eq. (\ref{scali}), augmented
by the relevant scalings of the matrices and of the twist-phases,
drastically changing the large-$N$ behavior of the two-matrix
model: in particular new non-trivial classical solutions could
emerge from a saddle-point analysis. Before closing the
section we have to mention that in \cite{nisj} eq. (\ref{scala})
was assumed, justifying the choice by the request that Wilson
loops much smaller then the noncommutativity scale agrees with the
commutative planar theory. The basic assumption there was
that Wilson loops at $\theta\to\infty$ reduce to the
usual ones of large-$N$ $2D$ Yang-Mills. We know that there are
evidences \cite{bacc}, derived from perturbative computations,
that this may be not true.

\noindent In order to confirm our expectations, we are going to
study the classical solutions of the TEK model, in the double
scaling limit relevant for the noncommutative theory: we assume
that eq. (\ref{scali}) is valid and we will find the appearance of
non-trivial configurations, potentially changing the structure of
the weak-coupling regime of conventional two-dimensional lattice
gauge theories.

\section{Equations of motion and their solutions}
In this section we solve the equations of motion for the TEK model
in two dimensions: we can treat the problem in full generality,
without referring to the particular dimension of the matrices or
to the peculiar structure of the twists. The specific properties
of the different models as well as the scaling of $\beta$ in
taking the continuum limit will be discussed in the next section.
We start by considering the general action
\begin{equation}
S_{TEK}=\beta N{\rm Tr}\Bigl[\Bigl(e^{-\pi i m/n}V_1 V_2-e^{\pi i m/n}V_2
V_1\Bigr) \Bigl(e^{i\pi m/n}V_2^\dagger V_1^\dagger-e^{-\pi i
m/n}V_1^\dagger V_2^\dagger\Bigr)\Bigr], \label{tek4}
\end{equation}
$V_1,V_2$ being unitary $n\times n$ matrices and $m,n$ a couple of
integers, that will be taken relatively prime. Eq. (\ref{tek4})
makes also manifest the positive nature of $S_{TEK}$.

\noindent The model enjoys the gauge symmetry
\begin{equation}
V_1\rightarrow C^\dagger V_1 C,  \ \ \ \
{\rm and} \ \ \ \ V_2\rightarrow C^\dagger V_2 C,
\end{equation}
with $C$ a unitary matrix: of course $V_1$ and $V_2$ do not
transform under the $U(1)$ subgroup, the effective symmetry group
being therefore $SU(n)/\mathbb{Z}_n$. The classical vacuum
solution (namely the absolute minimum of the action) is determined
by the condition
\begin{equation}
\label{minima1}
e^{-\pi i m/n} V_1^0 V_2^0-e^{\pi i m/n} V_2^0
V_1^0=0,
\end{equation}
that can be solved in terms of twist-eaters (see eqs.
(\ref{eaters}) and (\ref{eaters1})): it produces an irreducible
representation of the two-dimensional Weyl-'t Hooft algebra,
since ${\rm gcd}(m,n)$ is taken to be $1$. The gauge inequivalent solutions of
eq. (\ref{minima1}) are labelled by the global $U(1)$ phases
multiplying $V_1^0,V_2^0$. We can therefore associate to the
absolute minimum of the action eq. (\ref{tek4}) a moduli space,
with the topology of a torus, described by a pair of complex
moduli $(z_1,z_2)\in \tilde{T}^2$.

\noindent The equations of motion defining all the other extrema
are easily obtained by differentiating  the action with respect to
$V_1$ and $V_2$. We get
\begin{equation}
V_1^\dagger (W-W^\dagger)V_1-(W-W^\dagger)=0\ \ \ \
{\rm and }\ \ \ \
V_2^\dagger (W-W^\dagger)V_2-(W-W^\dagger)=0,
\end{equation}
where
\begin{equation}
W=e^{-2\pi i m/n}V_1 V_2 V_1^\dagger V_2^\dagger.
\end{equation}
These two equations possess a simple geometrical interpretation: the difference $(W-W^\dagger)$, which in the
large-$N$ limit is related to the field strength of the noncommutative theory, is covariantly constant when we move
in the direction $1$ or in the direction $2$. In the matrix language the above equations simply assert that the
difference $(W-W^\dagger)$ commutes with the unitary matrices $V_a,\,a=1,2$.

\noindent
To find the general solutions of these equations we have found useful to use the projectors
technique: the first step is to introduce the spectral decomposition of the matrix $W$
\begin{equation}
W=\sum_{j} e^{i\phi_j} P_j.
\end{equation}
Here $P_j$ is the orthogonal projector on the $j-$th eigenspace of $
W$ and $\exp(i\phi_j)$ is its eigenvalue. They satisfy the properties
\begin{equation}
\label{eq10}
P_k P_j=\delta_{k j} P_j \ \ \ \
{\rm and}
\ \ \ \
\sum_j P_j=\mathbb{I}.
\end{equation}
We recall that  this decomposition exists since this matrix is unitary.
In the following, however, the relevant combination is only the difference
$W-W^\dagger$, whose spectral decomposition is then
\begin{equation}
\label{spec1}
W-W^\dagger=2 i\sum_{j} \sin\phi_j P_j.
\end{equation}
We must notice that eigenspaces corresponding to different eigenvalues for
the matrix $W$ can merge and correspond to the same eigenvalue for the
above difference. This occurs when the matrix $W$ possesses both the
eigenvalue $\exp(i\phi_j)$ and the eigenvalue $\exp(i(\pi-\phi_j))$.
In fact, it is easy to see that they produce the same eigenvalue for the
difference $W-W^\dagger$ and thus their eigenspaces coalesce. For this
reason, we shall rewrite eq. (\ref{spec1}) as the reduced sum
\begin{equation}
W-W^\dagger=2 i\sum_{\ell} \sin\phi_\ell \mathbb{P}_\ell,
\end{equation}
where $\mathbb{P}_\ell$ are the orthogonal projectors on the eigenspaces
of the anti-hermitian matrix $W-W^\dagger$ and they satisfy the analogous of
eq. (\ref{eq10}). We have also that $\mathbb{P}_\ell=P_\ell$ if
$\exp(i\phi_\ell)$ is  an eigenvalue of $W$, but $\exp(i(\pi-\phi_\ell))$
is not. If they are both eigenvalues and $P_{1\ell}$ and $P_{2\ell}$ are
 the corresponding projectors,  $\mathbb{P}_\ell=P_{1\ell}+P_{2\ell}$.

\noindent
The equations of motion are now equivalent to the fact the matrices $V_a$
commute with the projectors $\PP_j$, namely
\begin{equation}
\PP_j V_a=V_a \PP_j.
\end{equation}
Let us now define the reduced matrices
\begin{equation}
\label{reduced}
\mathbb{V}^{(j)}_a=\PP_j V_a \PP_j,
\end{equation}
then along the equations of motion
\begin{equation}
V_a=\sum_j \mathbb{V}^{(j)}_a=\sum_j \PP_j V_a \PP_j.
\end{equation}
In other words, the matrices $V_a$ can be put in a block-diagonal form
since the eigenspaces defined by the projectors $\PP_j$ are invariant subspaces.
The reduced matrices $\mathbb{V}_a^{(j)}$ are also unitary on their eigenspace,
\begin{equation}
\mathbb{V}_a^{(j)^\dagger}\mathbb{V}_a^{(j)}
=\PP_j V_a^\dagger \PP_j^2 V_a \PP_j=
\PP_j V_a^\dagger V_a \PP_j^3=\PP_j^4=\PP_j.
\end{equation}
\noindent Since the matrices $V_a$ are block-diagonal the equations of motion
are trivially satisfied. Actually, we have still to impose that our spectral
decomposition holds or equivalently in each subspace  we must require
\begin{equation}
\PP_j( W-W^\dagger)\PP_j = 2 i \sin\phi_j \PP_j.
\end{equation}
This relation does not hold automatically because the matrix $W$ depends on
the matrices $V_a$ and it gives a constraint on the form of the matrices $V_a$
in each subspace. It can be rewritten  in terms of reduced matrices eq. (\ref{reduced}). Namely, if we define
\begin{equation}
\label{def}
\mathbb{W}^{(j)}=e^{-2\pi i m/n}\V_1^{(j)}\V_2^{(j)}\V_1^{(j)\dagger}
\V_2^{(j)\dagger}
\end{equation}
the above equation reads
\begin{equation}
\label{redeqs}
\mathbb{W}^{(j)}
-\mathbb{W}^{(j)\dagger} = 2 i\sin\phi_j \PP_j.
\end{equation}
At this point we simply have the two possibility already discussed:
\begin{itemize}
\item {\bf type I:}\\
 $\exp(i\phi_j)$ is an eigenvalue of $W$, but $\exp(i(\pi-\phi_j))$
is not. Then $\PP_j=P_j$ and we can write
\begin{equation}
\label{eqw1}
\mathbb{W}^{(j)}=
e^{-2\pi im/n}\V_1^{(j)}\V_2^{(j)}\V_1^{(j)\dagger}\V_2^{(j)\dagger}
=\exp(i\phi_j) P_j.
\end{equation}
The value of $\exp(i\phi_j)$ can be now determined by taking the determinant
of both sides. We have
\begin{equation}
\label{eqw12}
\exp({-2\pi i n_j\frac{m}{n}})
=\exp(i n_j\phi_j)\ \ \ \ \Rightarrow\ \ \ \ \phi_j=2 \pi (\frac{m_j}{n_j}-\frac{m}{n}),
\end{equation}
where $n_j$ is the dimension of the subspace and $m_j$ is an integer number
that runs from $0$ to $n_j-1$. Now eq. (\ref{eqw1}) on the restricted
subspace is
\begin{equation}
\label{eqw13}
\V_1^{(j)}\V_2^{(j)}
=\exp(2\pi i \frac{m_j}{n_j}) \V_2^{(j)}\V_1^{(j)}.
\end{equation}
This is exactly the equation for the twist-eaters of general twist, whose solutions are widely
discussed in the literature. We remark that, at this level, fixed the dimension $n_j$ of the subspace we have a different
solution for any choice of $m_j$: this is not the end of the story because $m_j$ and $n_j$ are not coprime,
in general, and therefore the space of the solutions needs a more refined treatment.
When gcd$(m_j,n_j)=\hat{n}_j\neq 1$, the representation of the Weyl-'t Hooft algebra eq. (\ref{eqw13}) is no longer
irreducible, having $\hat{n}_j$ irreducible components. It means that, up to gauge transformations, we can
always take $\V_1^{(j)},\V_2^{(j)}$ to be block diagonal, the $\hat{n}_j$ blocks being irreducible: we could expect
naively a solutions space
\begin{equation}
M_{(m_j,n_j)}=(\tilde{T}^{2})^{\hat{n}_j}=\tilde{T}^2\times\tilde{T}^2...\times\tilde{T}^2,
\label{moduli1}
\end{equation}
labelling the freedom in choosing the $U(1)$ phases of the
$\hat{n}_j$ blocks. We have instead to consider also the residual
gauge symmetry which acts by permuting the $\hat{n}_j$ irreducible
components: this action, on the solutions space eq.
(\ref{moduli1}), is represented by the permutation group
$S_{\hat n_j}$, and therefore the moduli space is the symmetric
orbifold \cite{cr}
\begin{equation}
{\cal M}_{(m_j,n_j)}={\rm
Sym}^{\hat n_j}\hat{T}^2=(\tilde{T}^2)^{\hat n_j}/S_{\hat n_j}. \label{moduli2}
\end{equation}
We have seen that the general solution of the equations of motions
is always block-diagonal: we can evaluate directly, therefore, the
contribution of the $j-$th block to the classical action. The
relevant quantity is ${\rm tr}(\mathbb{W}^{(j)})+{\rm
tr}(\mathbb{W}^{(j)\dagger})$ and, happily, it can be computed
without an explicit knowledge of the form of the solution. In fact
we have
\begin{equation}
{\rm tr}(\mathbb{W}^{(j)})+{\rm tr}(\mathbb{W}^{(j)\dagger})= 2
n_j \cos\phi_j=2 n_j \cos\left(2 \pi(\frac{m_j}{n_j}-\frac{m}{n})
\right).
\end{equation}
Let us notice that the explicit form of the original twist enters, in our procedure, directly at level of
classical action.
\item {\bf type II:}\\
Both $\exp(i\phi_j)$ and $\exp(i(\pi-\phi_j))$ are an eigenvalues of
$W$. Denoting with  $P_{1j}$ and $P_{2j}$ the corresponding projectors, we
have
\begin{equation}
\label{eqw2}
\mathbb{W}^{(j)}=
e^{-2\pi i m/n}\V_1^{(j)}\V_2^{(j)}\V_1^{(j)\dagger}\V_2^{(j)\dagger}
=\exp(i\phi_j) P_{1j}-\exp(-i\phi_j)P_{2j}.
\end{equation}
We have now to find two matrices that satisfies these equation. Again,
taking the determinant (on this subspace) of both sides, we get an
equation
\begin{equation}
\label{eqw21}
\exp({-2\pi i (d_{1j}+d_{2j})m/n})
=\exp(i\phi_j d_{1j} +id_{2j}(\pi-\phi_j)),
\end{equation}
that relates the parameters $\phi_j$ to the dimensions $d_{1j}$
and $d_{2j}$ of the subspaces. The existence of this kind of
extrema was noticed in ref. \cite{vanbaal} and it was not clear if
they could play some role in the large-$N$ limit: it was suggested
that they are the analog of multi-instantons configurations. We
will show, in the next section, that they are suppressed here as
$N$ becomes large. To this aim we have to first evaluate their
block-contributions to the classical action. Even in this case, the
sum ${\rm tr}(\mathbb{W}^{(j)})+{\rm tr}(\mathbb{W}^{(j)\dagger})$
can be easily computed and one finds
\begin{equation}
{\rm tr}(\mathbb{W}^{(j)})+{\rm tr}(\mathbb{W}^{(j)\dagger})= 2
(d_{1j}-d_{2j}) \cos\phi_j.
\end{equation}
\end{itemize}
\noindent The most general solution of the TEK model can be built,
therefore, by considering the direct sum of solutions of type I
and type II. It is quite clear that to any partition of $n$ we can
associate, in principle, a solution: be
$\{\nu_j\}=\{\nu_1,\nu_2,...,\nu_n\}$ the natural numbers
describing the partition
\begin{equation}
n=\nu_1+2\nu_2+..+n\nu_n.
\end{equation}
We can associate to $\{\nu_j\}$ a subspaces decomposition with $\nu_1$ one-dimensional subspaces ($d_1=1$),
$\nu_2$ two-dimensional subspaces ($d_2=2$) and so on. On any of these subspaces we can have or solutions of
type I or solutions of type II. For type I solutions we have been able to associate another
integer, $m_j$, allowing for an explicit realization in terms of twist-eaters , while for type II we did not succed
in giving such a simple description. This is not the end of the story, of course, we still have the freedom to weight
any subspace solutions with an arbitrary $U(1)$ phase: more
importantly we have to refine the relation between partitions and distinct solutions. In fact it is not difficult to
show that, for type I configurations, different choices of $(n_j,m_j)$ result in the same solution. We will discuss
the necessary refinement in the next section, in connection also with the moduli space structure of type I family.
 {}

\section{Solutions of Finite Action in the Double Scaling Limit}
In the previous section we have solved the classical equations of
motion for the two-dimensional TEK model: in order to recover the
noncommutative gauge theory we have still to perform the
double scaling limit. It is well known \cite{polly,andy,gn2} that
on the noncommutative plane finite action solutions exist and we
have shown in \cite{gsv} that the partition function and Wilson
lines correlators seems to be localized around them. More recently
\cite{pasz,pasz2} the same properties have been shown to hold even
at finite volume, where a localization theorem has been proven and
an explicit form of the partition function has been proposed.
We expect therefore that some of our TEK classical solutions
survive the relevant double scaling limits (i.e. their classical
action remain finite) and reproduce at least some instanton-like
feature.

\noindent
Let $\ell$ be the index running over the  solutions of type I and $n_\ell$ the corresponding
dimensions. In the same way let $s$ be the index spanning the solutions
of type II and $D_{1s}$ and $D_{2s}$ the dimensions of the associated
subspaces. Then the value of the action on a classical solution is
\begin{eqnarray}
&&S_{TEK}=\beta N\left(2 n-2 \sum_\ell n_\ell \cos\phi_\ell
-2\sum_s (D_{1s}-D_{2s}) \cos\phi_m\right).
\end{eqnarray}
Recalling the sum rule
\begin{equation}
\sum_\ell n_\ell+\sum_s (D_{1s}+D_{2s})=n,
\end{equation}
the above equation can be rewritten as a sum positive definite objects
\begin{eqnarray}
\label{eq3.3}
&&S_{TEK}=\beta N\left(\sum_\ell n_\ell
\sin^2\left(\frac{\phi_\ell}{2}
\right)
+\sum_s \left(D_{1s}\sin^2\left(\frac{\phi_s}{2}
\right)
+D_{2s} \cos^2\left(\frac{\phi_s}{2}
\right)\right)
\right).
\end{eqnarray}
In the following we shall look for solutions such that $S_{TEK}$ is
finite in the large-$N$ limit: we have in mind, of course, to send also $n$ to infinity, according to
the precise scalings derived in Sect. 3. Since the overall coefficient in eq. (\ref{eq3.3}) 
diverges in this limit, both in recovering the NC torus and the NC plane, 
this occurs only when each term in the quantity between parenthesis goes to zero quickly enough. Namely, we must have
\begin{equation}
\label{eq3.4}
n_\ell \sin^2\left(\frac{\phi_\ell}{2}
\right)\to 0
\end{equation}
and
\begin{equation}
\label{eq3.5}
D_{1s}\sin^2\left(\frac{\phi_s}{2}
\right)\to 0
\ \ \ \ \ \
D_{2s} \cos^2\left(\frac{\phi_s}{2}
\right)\to 0.
\end{equation}
First, we examine eq. (\ref{eq3.5}): if both $D_{1s}$ and $D_{2s}$ are
different from zero, this condition cannot be satisfied. In fact it would
require that the sine and the cosine of the same angle vanish. If one of
the two dimensions is zero, the solution of type II collapses in a solution
of type I. This case is impossible because we assume to have already
counted all the solutions of type I in the first sum. So the only possibility
left is
\begin{equation}
D_{1s}=D_{2s}=0\ \ \  \ \ \forall\  s.
\end{equation}
In other words, solutions of type II have not finite action in the large-$N$ limit:
we are left with the solution of type I and the value of $S_{TEK}$ is
\begin{equation}
\label{eq3.7}
S_{TEK}=\beta N\sum_\ell n_\ell
\sin^2 \pi \left (\frac{m_\ell}{n_\ell}-\frac{m}{n}\right),
\end{equation}
where
\begin{equation}
\sum_\ell n_\ell=n\  \ \ \ \ {\rm and}\  \ \ \ \ m_{\ell}=0,\dots,n_\ell-1.
\end{equation}
In order to have finite limit, each term in the sum eq. (\ref{eq3.7}) must be
finite, since no cancellation can intervene among different terms, being in fact all positive.
We start by discussing the noncommutative torus.

\subsection{Solutions with finite action on the torus}

Let us consider the general case with $p$ and $q$ coprime: $m/n$ is given by eq. (\ref{msun}) and
the dimension of the matrices is $Np-2rq$. We have to impose the scaling on $\beta$: as we
have already discussed, we choose
\begin{equation}
\beta=\frac{2}{g^2 a^2}=\frac{2N^2}{g^2A},
\label{scalata}
\end{equation}
to perform the double scaling limit. In fact, the coupling scales canonically
with the lattice spacing $a$, namely $\beta\propto a^{-2}\propto N^2$. Here $A$ denotes the
area of the noncommutative torus and $g^2$ is the physical coupling constant, the factor 2 being
inserted for later convenience. Only the type I solutions can have a finite limit and the value of the action is
\begin{equation}
\label{eqqq3.7}
S_{TEK}=\frac{2}{g^2A}N^3\sum_\ell n_\ell
\sin^2\left(\frac{\phi_\ell}{2}
\right)= \frac{2}{g^2A}{N^3}\sum_\ell n_\ell
\sin^2 \pi \left (\frac{m_\ell}{n_\ell}-\frac{m}{n}\right),
\end{equation}
where
\begin{equation}
\sum_\ell n_\ell=Np-2rq\  \ \ \ \ {\rm and}\  \ \ \ \ m_{\ell}=0,\dots,n_\ell-1.
\end{equation}
To perform the limit we use the parametrization
\begin{equation}
n_\ell=N p_\ell- (2 r) q_\ell
\ \ \ \ \ \  m_\ell=-s p_\ell+k q_\ell,
\label{param}
\end{equation}
to rewrite the action in the form
\begin{eqnarray}
\label{eqqq3.8}
S_{TEK}&=&\frac{2}{g^2A} {N^4}\sum_\ell (p_\ell-\theta q_\ell)
\sin^2 \pi \left (\frac{-s p_\ell+k q_\ell}{N p_\ell- (2 r) q_\ell }
-\frac{-s p +k q}{N p- (2 r) q }\right)=\nonumber\\
&=&\frac{2}{g^2A} {N^4}\sum_\ell (p_\ell-\theta q_\ell)
\sin^2 \frac{\pi}{N^2} \left (\frac{q_\ell}{p_\ell- \theta q_\ell }
-\frac{q}{ p- \theta q  }\right).
\end{eqnarray}
At this point we take the large-$N$ limit, assuming that $\theta=2r/N$ approaches
an irrational number: we easily get
\begin{eqnarray}
\label{fippo}
S_{TEK}
=\frac{2\pi^2}{g^2A} \sum_\ell (p_\ell-\theta q_\ell)
                       \left (\frac{q_\ell}{ p_\ell- \theta q_\ell }
-\frac{q}{p- \theta q }\right)^2.
\end{eqnarray}
This is the central result of the paper: eq. (\ref{fippo}) exactly reproduces the value
of the action of Yang-Mills theory on a $(p,q)$ projective module, evaluated on the classical solutions
of the equations of motion \cite{pasz,cr}. To get a complete identification we have to specify the range
of $p_\ell,q_\ell$ and to show they satisfy indeed the correct constraints.
\noindent
The first thing to notice is that the positivity of the dimensions implies
\begin{eqnarray}
Np-2rq>0\,\,\,\,\, &\Rightarrow&\,\,\,\,\, p-\theta q>0\nonumber\\
n_\ell >0\,\,\,\,\, &\Rightarrow&\,\,\,\,\, p_\ell-\theta q_\ell >0.
\label{positivi}
\end{eqnarray}
Then we have to establish a sum rule over $p_\ell$ and $q_\ell$:
one would be tempted to set the sums equal to $p$ and $q$, namely
\begin{equation}
\sum_\ell q_\ell=q\ \ \ \ \ \ \
\sum_\ell p_\ell=p.
\end{equation}
This cannot be done in a straightforward manner. On $p_\ell$ and $q_\ell$
we have only the constraint $\sum_\ell n_\ell=Np-2rq$, which in turn implies
\begin{equation}
\sum_\ell q_\ell=q-N j\ \ \ \ \ \ \
\sum_\ell p_\ell=p-2 r j .
\end{equation}
However if $2r/N$ approaches an irrational number the only consistent
choice  in the large-$N$ limit is $j=0$. This gives the desired sum
rules for $p_\ell$ and $q_\ell$, establishing the connection with the "partitions"
described in \cite{pasz,cr}. The total moduli space is also seen to be trivially the same, in the
limit we are considering: following \cite{pasz}, a given partition $(p_\ell,q_\ell)$,
compatible with eq. (\ref{positivi}), can be unambiguously presented as $(N_\ell',p_\ell',q_\ell')$
with $p_\ell',q_\ell'$ relatively prime and $N_\ell'$ is the associated multiplicity.
From eq. (\ref{param}) we obtain, at level of TEK solutions, the related presentation as
$(N_\ell',m_\ell',n_\ell')$: the integers $N_\ell'$ simply count the number of irreducible representations of a given
Weyl-'t Hooft algebras (labelled by the relatively prime integers $(m_\ell',n_\ell')$) inside a given solution and
they are enough to construct the total moduli space
\begin{equation}
{\cal M}_{(m,n)}=\Pi_\ell{\cal M}_{(N_\ell' m_\ell',N_\ell' n_\ell')}=\Pi_\ell{\rm Sym}^{N_\ell'}\tilde{T}^2,
\label{modulitot}
\end{equation}
the $\ell$ index running over the same partitions defined by eq. (\ref{positivi}). Once the relation between
$(N_\ell',n_\ell',m_\ell')$ and $(N_\ell',p_\ell',q_\ell')$, as determined by eq. (\ref{param}) in the large-$N$ limit,
is realized, we have the full identification of the moduli space of classical solutions of Yang-Mills theory on
the $(p,q)$ projective module with the moduli space of our finite action TEK classical solution
\begin{equation}
{\cal M}_{(p,q)}={\cal M}_{(m,n)}.
\end{equation}

\noindent
The emerging of these finite action configurations crucially relies on the scaling eq. (\ref{scalata}) and
on the judicious subtractions performed in defining the starting TEK action. The first one, appearing in
eq. (\ref{sub}), makes the action always positive and it corresponds to the subtraction of a divergent term,
in the large-$N$ limit, independent of the details of the particular projective module. It is related to
the zero-point energy of the system and it usually appears also in conventional lattice theory. The second
subtraction is subtler and it is performed by fixing the background connection in such a way that the absolute
minimum of the action, in a given $(p,q)$ sector, is at zero value, for the constant curvature connection characterizing
the projective module itself. This choice produces a delicate cancellation between the difference inside
the sine in eq. (\ref{eqqq3.8}), resulting into the $1/N^2$ decaying of the argument instead of a generic
$1/N$. Conversely, after taking the limit, we obtain the classical action in presence of a background connection,
having minimum zero at the correct value (see \cite{pasz} for a discussion of the subtraction on the
continuum). It could be nevertheless useful to choose a different value of the continuum background connection:
in particular, we can fix the minimum in such a way that eq. (\ref{fippo}) reproduces the commutative value of the action,
for a given Chern class, in the limit $\theta\to 0$. We modify therefore eq. (\ref{connec}) as
\begin{equation}
\phi\to\phi+i\pi \frac{q}{nN}=\phi+\frac{i\pi}{N^2}\frac{q}{p-\theta q}:
\end{equation}
it is simple to verify that eq. (\ref{fippo}) changes as
\begin{equation}
\label{fippo1}
S_{TEK}
=\frac{2\pi^2}{g^2A} \sum_\ell
                       \left (\frac{q_\ell^2}{ p_\ell- \theta q_\ell }\right).
\end{equation}
Let us notice that the modification we have done is of order $1/N^2$, preserving in this way the delicate
balance we have observed to produce the finite action solutions. Nicely, as $\theta\to 0$, we can recover
the classical configurations solving  the commutative theory: for $p=1$ we have the unique choice
$q_\ell=q$, giving us the $U(1)$-instanton associated to the Chern number $q$. For general $p$ we obtain the solutions
of the $U(p)$ theory in the relevant charge sector. Having recovered the classical solutions of the commutative
theory for finite rank of the gauge group from a one-plaquette  model is not trivial at all: in fact, as we will discuss
in Sect. 6, the exact solution of the familiar YM$_2$ involves a different kind of lattice discretization.

\noindent
We see that eq. (\ref{fippo}) confirms our interpretation of the parameter $q$ as the Chern class associated to the module.
However it would be interesting to have a direct computation of this quantity as the discretized integral of
the lattice field strength. The equations of motions and a simple analysis for small lattice spacing suggest
to identify the lattice field strength with
\begin{equation}
F=\frac{i}{2}\Bigl(D_1 D_2 D_1^\dagger D_2^\dagger-D_2 D_1 D_2^\dagger D_1^\dagger\Bigr),
\end{equation}
in the large-$N$ limit. Then its trace (the integral in the noncommutative language) produces
\begin{equation}
Q=\frac{i}{2}{\rm Tr}\Bigl[D_1 D_2 D_1^\dagger D_2^\dagger-D_2 D_1 D_2^\dagger D_1^\dagger\Bigr].
\end{equation}
We have
\begin{eqnarray}
Q&=&\frac{i}{2}N\sum_\ell  \Bigl[n_\ell\exp(2\pi i s/N)
\exp(2\pi i (k q_\ell-p_\ell s)/n_\ell)-{\rm c.c.}\Bigr]
\nonumber\\
&=& N^2\sum_\ell (p_\ell-\theta q_\ell)\sin\left(
\frac{2\pi q_\ell}{N^2(p_\ell-q_\ell\theta)}    \right),
\end{eqnarray}
that in the large-$N$ limit gives us the desired relation
\begin{equation}
Q=2\pi  \sum_\ell q_\ell=2\pi q.
\end{equation}
\subsection{Solutions with finite action on the plane}
Reaching the noncommutative plane  is actually  tricker than the torus. We recall,
in fact, that the only relevant finite quantity in the large-$N$ limit is  the
dimensional combination
\begin{equation}
\Theta_{plane}=\frac {\theta A}{2\pi}= N^2 a^2 \frac{2 r}{2N\pi} = \frac{r N a^2}{\pi}.
\end{equation}
Its finiteness, in turn, implies that we can drop the dependence on the area in all our
expressions in favor of that on $\Theta_{plane}$,
\begin{equation}
A= \frac{\pi N \Theta_{plane}}{r }.
\end{equation}
Then the value of the  $S_{TEK}$ action eq. (\ref{eqqq3.8}) is more conveniently written as
\begin{equation}
\label{ciap}
S_{TEK}=\frac{2r}{\pi g^2 \Theta_{plane}} {N^3}\sum_\ell (p_\ell-\theta q_\ell)
\sin^2 \frac{\pi}{N^2} \left (\frac{q_\ell}{p_\ell- \theta q_\ell }
-\frac{q}{ p- \theta q  }\right).
\end{equation}
Two remarks immediately stem from a first inspection of eq. (\ref{ciap}): the factor in front
of the action scales as $N^3$  (and not as $N^4$)  and the parameter $\theta$ is no longer a
finite quantity in the large-$N$ limit, but its values flows to zero as $1/N$. Thus, differently
from what happened in the torus case,  each term in the sum   behaves as $1/N$
and it  possesses a vanishing limit
\begin{eqnarray}
&&\!\!\!\!
\frac{2r{N^3}(p_\ell-\theta q_\ell)}{\pi g^2 \Theta_{plane}}
\sin^2 \frac{\pi}{N^2} \left (\frac{q_\ell}{p_\ell- \theta q_\ell }
-\frac{q}{ p- \theta q  }\right)\simeq\frac{2r \pi^2 {N^3}(p_\ell-\theta q_\ell)}{\pi g^2 \Theta_{plane}N^4}
  \left (\frac{q_\ell}{p_\ell- \theta q_\ell }
-\frac{q}{ p- \theta q  }\right)^2\!\!\!\!\!
\simeq\nonumber\\
&&\!\!\!\!
\simeq\frac{2\pi r (p_\ell-\theta q_\ell)}{ g^2 \Theta_{plane}N }
 \left (\frac{q_\ell}{p_\ell- \theta q_\ell }
-\frac{q}{ p- \theta q  }\right)^2\longrightarrow 0.
\end{eqnarray}
The classical solutions of the TEK model in the large-$N$ limit defining the plane appear to form a
democratic sea of vanishing action configurations. No sign of the fluxons discussed in \cite{polly,gn2}
is apparently present.

\noindent
Fortunately, this way of reasoning has a welcome exception when some of the $p_l$ in the partition of
$p$ vanish.  When $N\to\infty$, the corresponding term in the action takes the form
\begin{eqnarray}
&&\!\!\!\!
\frac{2r{N^3}(-\theta q_\ell)}{\pi g^2 \Theta_{plane}}
\sin^2 \frac{\pi}{N^2} \left (\frac{1}{\theta}
+\frac{q}{ p- \theta q  }\right)\simeq\frac{2r {N^3}(-\theta q_\ell)}{\pi g^2 \Theta_{plane}}
 \frac{\pi^2}{N^4\theta^2} \left (1
+\frac{q\theta}{ p- \theta q  }\right)^2\!\!\!\!\!
\simeq\nonumber\\
&&\!\!\!\!
\simeq-\frac{2\pi r  q_\ell}{ g^2 \Theta_{plane}N \theta}
 \left (1
+\frac{q\theta}{ p- \theta q  }\right)^2\longrightarrow -\frac{ \pi q_\ell}{g^2 \Theta_{plane}},
\end{eqnarray}
where we have used that $N\theta=2 r$.  The limit is totally independent of the free parameter
$r$ that we had at the beginning in the  definition of $\Theta_{plane}$. Collecting the  different
contributions, the total action is then
\begin{equation}
S_{plane}= -\sum_{\hat\ell}\frac{ \pi q_{\hat\ell}}{ g^2 \Theta_{plane}}=-\frac{\pi\sum_{\hat\ell} q_{\hat\ell}
}{g^2 \Theta_{plane}}
\equiv -\frac{ \pi\hat k
}{g^2 \Theta_{plane}},
\end{equation}
where the sum runs only over the elements of the partition with vanishing $p_\ell$. This is exactly
the value  of the action for the fluxons discovered by \cite{polly,gn2}. In fact it exhibits the two
peculiar features of these topological objects, solving the classical  Yang-Mills equations
on the noncommutative plane: its value grows linearly with the topological charge at variance with quadratic behavior
for the instanton solutions on the torus. Moreover the sign of total charge $\hat k$ obeys the positivity
constraint
\begin{equation}
-\Theta_{plane} \hat k>0,
\end{equation}
which simply asserts that the sum of the dimensions of the eigenspaces contributing is positive. This also
displays the chiral nature of these solutions.

\noindent
The picture emerging from the above limiting procedure deserves some comments. First of all,
we notice that the same value of the action can be obtained  starting from very different configurations
at the level of the matrix model: the only necessary ingredient is the presence of a certain number of
vanishing $p_l$ whose corresponding $q_\ell$ sum to the desired total charge $\hat k$. The possible
choices of the relevant $q_\ell$ are therefore in correspondence with the partitions of $|\hat k|$. This
inner structure should have a natural interpretation in terms of the moduli space of fluxons  \cite{ssg}.

\noindent
There is another subtle source of degeneracy, already announced   at the beginning of this section:
the subset of the partition with non vanishing $p_\ell$ is more or less unconstrained, since
it gives a vanishing contribution to the total action. The only conditions that the $p_\ell$ must
fulfill are
\begin{equation}
\sum_{\ell} p_\ell=p \ \ \ \ \mathrm{with} \ \ \ p_\ell>0,
\end{equation}
while the corresponding $q_\ell$ sum to $q-\hat k$. The positivity of the $p_\ell$ is what
survives of eq. (\ref{positivi}) in the large-$N$ limit. It is worth noticing that  the additional
structure  generated by the $p_\ell$ strongly resembles that of a commutative  instanton of
$U(p)$ with Chern class equal to $q-\hat k$ on the torus and thus it seems to carry all the original
geometrical data. It is also well-known that these commutative configurations
become  degenerate with the vacuum in the decompactification limit, as it happens here.
Notice that the above degeneracy disappear for $p=1$ and the large area limit can be
carried in straightforward manner safely reaching the so-called $U(1)$ Yang-Mills theory
on the noncommutative plane. This was also the choice made in ref. \cite{gsv}.

\noindent
The exact correspondence between fluxons with their ancestors on the noncommutative
torus as well as their contribution to the partition function on the plane will be
discussed in details in ref.  \cite{ssg}.

\section{A few remarks on the quantum theory}
The success in describing the instantons over the noncommutative torus
strongly suggests that a complete quantum analysis of NCYM$_2$ may be not out of reach in our
framework. This hope is also corroborated by the
observation that NCYM$_2$ is expected to be  semiclassically exact \cite{pasz}: namely the quantum
observables can be expressed as a sum over the classical solutions, weighted with a fluctuations factor.

\noindent
However the situation is more intricate than one can naively think. The path that has
led Migdal to solve the usual QCD$_2$ encounters here a couple of serious obstacles: the continuum
limit is, as already stated many times, intrinsically tied  with a  large-$N$ limit, while in the
ordinary theory it can be safely taken at finite $N$. Second, a group theory characterization
of the emerging theory is quite difficult, the representation theory of $U(\infty)$ being almost unknown
at variance with that of $U(N)$.

\noindent
Let us start by discussing some general facts: the computation of the partition function,
\begin{equation}
\mathcal{Z}_{TEK}\!\!=\!\!\ \int\!\!\! \frac{D V_1 D V_2}{\mathrm{Vol}(U(n))} \exp\Biggr( \beta N{\rm Tr}\biggr[
e^{2i\pi \frac{ m}{n}}V_1 V_2 V_1^\dagger V_2^\dagger+e^{-2i\pi \frac{ m}{n}}
V_2 V_1 V^\dagger_2 V_1^\dagger\biggl]-2\beta Nn\Biggl)
\end{equation}
can be reduced to a one-matrix integral through two steps. To show this we first introduce the following representation
of delta function over the unitary group in terms of the characters $\chi_R$
\begin{equation}
\sum_R \chi_R(V_1 V_2 V_1^\dagger V_2^\dagger)\chi_R(W^\dagger)=\delta(V_1 V_2 V_1^\dagger V_2^\dagger,W),
\end{equation}
with the sum running over all representations of $U(n)$, and then we perform the integral over $V_i$ by
means of the formula
\begin{equation}
\int\, DU\, {\chi}_R(UAU^\dagger B)=\frac{ \chi_R(A)\chi_R(B) }{d_R} .
\label{app21}
\end{equation}
The final result is
\begin{equation}
\mathcal{Z}_{TEK}\!\!=\!\!\ \int \frac{D W}{\mathrm{Vol}(U(n))}\exp\Biggr( \beta N
\left[ {\rm Tr}(W)+
{\rm Tr}(W^\dagger)\right]-2\beta Nn\Biggl)
\sum_R \frac{\chi_R(e^{-2i\pi \frac{ m}{n}}W^\dagger)}{d_R}.
\end{equation}
We can even compute the integral over $W$ using the formula
\begin{equation}
\lambda_R(\beta N)=\frac{1}{d_R}\int \frac{D W}{\mathrm{Vol}(U(n))} \exp\Biggr( \beta N
\left[ {\rm Tr}(W)+
{\rm Tr}(W^\dagger)\right]\Biggl)
\chi_R(W)
\end{equation}
where $\lambda_R(\beta N)$ can be explicitly expressed through the determinant of a matrix having Bessel functions
as entries \cite{dz}, whose form will be not relevant for our discussion. 
Accordingly the partition function takes the compact form
\begin{equation}
\mathcal{Z}_{TEK}=\sum_R  \lambda_R\left(\frac{2 N^3}{g^2 A}\right)
\exp\left({-2\pi i n_R \frac{ m}{n}-\frac{4 N^3 n}{g^2 A}}\right ),
\label{tek5}
\end{equation}
where $n_R$ is the total number of boxes in the Young tableaux associated the representation $R$
and we have also used that $\beta=2 N^2/g^2A$ for the noncommutative torus.
The other term in the exponential is, of course, the subtraction introduced in Sect. 2.  The apparent
simplicity of this expression is misleading,  hiding the complexity of the double scaling limit
necessary to reach the continuum. To better understand the roots of the difficulties that
one would encounter as $N\to \infty$, it is instructive to contrast this expression with the
usual QCD$_2$ on the lattice
\begin{equation}
\mathcal{Z}_{QCD_2}=\sum_R \left(\lambda_R\left( \frac{2 s^2}{g^2 A}\right)\right)^{s^2}
\exp\left({-4 N s^4/g^2 A}\right ),
\label{migda1}
\end{equation}
where $s^2$ is the total number of the sites. We recall that here $N$ has to be
kept fixed while $s$ goes to infinity. In this limit the combination
$$
\left(\lambda_R\left( \frac{2 s^2}{g^2 A}\right)\exp\left({-4 N s^2/g^2 A}\right )\right)^{s^2}
$$
can be nicely evaluated by using the asymptotic expansion of the Bessel functions, appearing in the
explicit expression of $\lambda_R$: up to an irrelevant multiplicative overall constant,
the leading contribution exponentiates producing the celebrated result \cite{mig}
\begin{equation}
\mathcal{Z}_{QCD_2}=\sum_R\exp\left(-\frac{g^2 A}{2}C_2(R)\right),
\end{equation}
$C_2(R)$ being the quadratic Casimir of the representation $R$. In spite of their superficial similarity,
there are both  technical and conceptual differences between eq. (\ref{tek5}) and eq. (\ref{migda1}).
For example, in the second case, the sum  over group representation
is a spectator in the continuum limit (large $s$), which in turn
can be taken term by term. Viceversa  in the first case, where
the continuum limit is at large $N$ (large group rank), the number of integers you sum over keeps growing.
It is also difficult, in this case, to isolate the asymptotic behavior of each term in the sum, since the determinant of
Bessel functions defining $\lambda_R$ depends on $N$ both through the form of its matrix elements and through its
dimension. These different behaviors reflect deeply two alternative ways to deal with space-time discretization: the usual
approach, where the lattice is an entity independent of its matter content contrasted with the noncommutative
point of view, where space-time and gauge symmetries are indissolubly tied.

\noindent
An attempt to investigate in a concrete way the large-$N$ limit of $\mathcal{Z}_{TEK}$
was performed in \cite{pasz3}, where the partition function has been
expressed through an expansion in inverse powers of the coupling constant: unfortunately the
authors have not been able to discuss the complexity of the double scaling limit. We shall
not try to tackle here this difficult problem, which would require a deeper understanding on how
classical configuration  should dominate the large-$N$ limit: recall, in fact, that the theory is
believed to be semiclassically exact.  However a simple computation, checking the consistency of the TEK approach
at quantum level, can be performed in a quite straightforward manner.
Let us consider in fact the following observables
\begin{equation}
\mathcal{O}_k=\frac{1}{n{\cal Z}_{TEK}}\int \frac {D V_1\,DV_2}{\mathrm{Vol}(U(n))}
\exp(-S_{TEK}){\rm Tr}\Bigl[\Bigl(V_1V_2V_1^{\dagger}V_2^{\dagger}\Bigr)^k\Bigr].
\label{app1}
\end{equation}
It is not difficult to show that $O_k$ is the quantum
average of a Wilson loop winding $k$-times around the fundamental plaquette, the area surrounded by the
contour being $a^2$ (we recall that $a$ is the lattice spacing). We want to compute this object in a
particular limit: when setting $a\to 0$, to reach the continuum limit, we also consider the winding $k$
very large so that
\begin{equation}
k^2 a^2=\lambda.
\label{app4}
\end{equation}
Following closely what we have done above, we can reduce the computation to the one-matrix integral
\begin{eqnarray}
&&\!\!\!\!\!\mathcal{O}_k=\\
&&\!\!\!\!\!\frac{1}{n{\cal Z}_{TEK}}
\int \frac{D W}{\mathrm{Vol}(U(n))}\exp\Biggr( \beta N
\left[ {\rm Tr}(W)+
{\rm Tr}(W^\dagger)\right]-2\beta Nn\Biggl){\rm Tr}[W^k]
\sum_R \frac{\chi_R(e^{-2i\pi \frac{ m}{n}}W^\dagger)}{d_R}.\nonumber
\label{app3}
\end{eqnarray}
In the following we shall restrict our investigation to the two-dimensional noncommutative plane (we
choose for simplicity $r=1$ and $N=n$):
$\mathcal{O}_k$,
in this case, corresponds to a Wilson loop average of vanishing area, winding an infinite number of times.
On the commutative plane observables of this type have been considered in \cite{bgv}: there it has been observed,
starting from the exact solution at finite $k$ and non-vanishing area, that this particular limit correspond to a truly
perturbative situation, being completely determined by the zero-instanton approximation on compact surfaces.
We expect, therefore, that also in the noncommutative case these observables can be captured by a perturbative computation.

\noindent
By exploiting the double scaling limit relevant for the plane we obtain a simple
relation between $k$ and $n$ in the form
\begin{equation}
k^2=\frac{\pi\lambda}{\Theta} n.
\label{app5}
\end{equation}
Let us now evaluate eq. (\ref{app3}) in the limit of large $k$: we expect, in this case, that the
integral is dominated by the classical solution $W=1\!\!1$. We write
\begin{equation}
W=\exp(i \hat{H})
\label{app6}
\end{equation}
where $\hat{H}$ is an hermitian matrix: we see that the rapid oscillations of the holonomy factor are controlled
by redefining $\hat{H}=H/k$, forcing therefore an expansion around $W=1\!\!1$. Within this approximation $O_k$ can
be expressed as an integral over the hermitian matrix $H$
\begin{equation}
\mathcal{O}_k=\frac{1}{{\cal Z}_{TEK}n}\sum_R \frac{e^{-2i\pi n_R\frac{ m}{n}}}{k^{n^2/2}}
\Biggl[\int\,\frac{D H}{\mathrm{Vol}(U(n))}\exp\Bigl(-\frac{\beta n}{k^2}{\rm Tr}[H^2]+i H\Bigr)+o(1/k)\Biggr].
\label{app62}
\end{equation}
The combination $\beta n/k^2$ turns out to be
\begin{equation}
\frac{\beta n}{k^2}=\frac{2n}{g^2\lambda}.
\label{app7}
\end{equation}
By computing ${\cal Z}_{TEK}$ in the same approximation and then using standard technique to evaluate
the matrix integral we arrive at
\begin{equation}
\mathcal{O}_k=W_n(\lambda)=\frac{1}{n}\exp(-\frac{g^2\lambda}{4n})L^1_{n-1}(\frac{g^2\lambda}{8n}),
\label{app8}
\end{equation}
expressed through the Laguerre polynomial $L^1_m(z)$. The limit $n\to\infty$ can be now easily obtained
\begin{equation}
W_n(\lambda)\to 2\sqrt{\frac{2}{\lambda g^2}}J_1(\sqrt{\frac{\lambda g^2}{2}}),
\label{app81}
\end{equation}
$J_1(z)$ being a Bessel function. The above result deserves some considerations: first of all the computation
we did it is essentially "perturbative", although justified by the large $k$-behavior. We have, in fact,
 perturbed the path-integral around the classical solution $W=1\!\!1$, that should correspond to the usual perturbative
expansion on the noncommutative plane. Secondly, we notice that the final result is independent from
the effective $\Theta$ parameter on the plane, suggesting that it has to be related to some kind of
planar limit. These properties can be tested by familiar field theoretical perturbative computations on
$\mathbb{R}_{\theta}$, using the star-product formalism: in \cite{bacc,bnt2} Wilson loops on noncommutative
plane were studied by perturbative methods. In particular the authors considered in \cite{bnt2} a $k$-winding circular 
contour, obtaining the complete ${\cal O}(g^4)$ result: it is not difficult to show that, in our limit, 
only the leading planar contribution survives in their expression, consistently reproducing eq. (\ref{app81}).

\section{Conclusions and outlook}
The TEK model naturally provides a non-perturbative definition of noncommutative Yang-Mills theory in
every dimension, which can be employed, in principle, for concrete computer simulations. In two dimensions
this approach becomes also a feasible tool for analytic computations due to the relative simplicity of
the relevant matrix model. Previous investigations in the continuum theory offer a number of results
to be checked and suggest the possibility to use the discrete formulation to unveil new
properties. We think, here, we have made a couple of steps in these directions: on one hand we have proposed
an unconstrained TEK model describing discretized noncommutative two dimensional Yang-Mills theory on a $(p,q)$
projective module. On the other, we have been able to solve the related classical equations of motion; more importantly
we have found a precise match between the solutions with finite action, in the relevant double scaling limit,
and the critical points of the continuum theory. The classical solutions on a
$(p,q)$ projective module possess a rather non-trivial structure: we hope that having recovered it from the
discretized formulation be a strong evidence of the correctness of our unconstrained approach. This encourages us
to take a further step and to try to solve completely the quantum theory within this framework. In doing so,
we could provide the natural derivation of the beautiful partition function proposed in \cite{pasz} and open the
possibility to compute general observables in a matrix model language. The fact the final theory should be semiclassical
exact, as shown in \cite{pasz}, makes this attempt not completely hopeless, having in our case a complete
understanding of the classical solutions and of their moduli space. The last section of the paper has been devoted to
elucidate what are the new issues that arise in trying to tackle the computation, when considered in its full
complexity: we hope to report in the near future about progresses in this direction.

\section*{Acknowledgements}
A special thanks goes to Richard Szabo for illuminating and friendly discussions.

\noindent
\,

\noindent
\,
\appendix

\noindent
{\Large \bf Appendices}

\section{Morita equivalence on the fuzzy torus}
The Morita equivalence  is  a surprising symmetry when seen from a conventional field theoretical
point of view: it connects indeed  theories with different  gauge groups, different coupling constants
and living on different tori.  Physically it can be understood as coming from the T-duality possessed
by the stringy ancestors  of these models, while mathematically it expresses the fact that certain classes of
algebras share the same representation theory.

\noindent
In particular in two dimensions the Morita equivalence transformations are realized by a group element of
$SL(2,\mathbb{Z})$,
\begin{equation}
\label{cippino7}
\left (
\matrix{a & b\cr c & d}\right).
\end{equation}
On the parameters, specifying the module, the transformation acts as
\begin{equation}
\label{cippino6}
\left(\matrix{p^\prime \cr q^\prime }\right)=
\left (
\matrix{a & b\cr c & d}\right)\left(\matrix{p\cr q}\right).
\end{equation}
In the commutative language, eqs. (\ref{cippino6}) corresponds to a non-trivial
change of the gauge group and the Chern class of the original theory. The noncommutative
parameters $\theta$ transforms with the M\"obius transformation associated to eq. (\ref{cippino7})
\begin{equation}
\label{cippino9}
\theta^{\prime}=\frac{ a\ \theta+b}{c\ \theta +d},
\end{equation}
while the dimension of the module $\mathcal{E}$ scales as
\begin{equation}
\mathrm{dim}\ \mathcal{E}^\prime=\frac{\mathrm{dim}\ \mathcal{E}}{|c\ \theta+d|}.
\end{equation}
Finally the invariance of the noncommutative Yang-Mills action dictates the transformation rules for
the remaining relevant objects
\begin{eqnarray}
L^\prime=|c\ \theta + d | L\ \ \ \ g^{\prime 2}=|c\ \theta+d| g^2\ \ \ \ \Phi^\prime=
\Phi (c \ \theta +d)^2-\frac{c(c\ \theta+d)}{2\pi R^2}.
\end{eqnarray}

\noindent
In the following  we will try to understand  how this symmetry is realized
in our description of the fuzzy torus. The first observation is that in
our framework the integer $N$ carries the information about the size $L=Na$
of the lattice and so its natural transformation under Morita is
\begin{equation}
N^\prime=N ( c \theta + d)= (2 c r+ d N).
\end{equation}
The transformation for $\theta$ eq. (\ref{cippino9}) can be now translated into a transformation
rule for integer $2 r$ appearing in our definition of the noncommutative lattice
\begin{equation}
2 r^\prime= (2 r a+b N).
\end{equation}
That the lhs of the above equation is even must be understood modulo $N^\prime$.
Here we have limited ourselves to the transformations for which the
combinations is $(2 c r+ d N)$ is positive. This is not really a
restriction, because it is equivalent to state that our duality group
is $PSL(2,\mathbb{Z})$.

\noindent
Now we stress that our final TEK model  is completely defined by three parameters
\begin{eqnarray}
n&=& N p - 2 r q,\nonumber\\
m&=&k q-p s,\nonumber\\
\beta N&=&\frac{2 N^3}{g^2 A},
\end{eqnarray}
the last one being the total coefficient in front of the  action.
Let us see how they change under Morita. For the first one, we have
\begin{eqnarray}
n^\prime&=& N^\prime p^\prime-2 r^\prime q^\prime=
(2 c r+ d N) (a p+b q)-(2 r a+b N)
(c p+d q)=\nonumber\\
&=&({2 a c r p} + 2  b c r q +a d N p +b d Nq
-{2 r a c p} - 2 r c b  N -b c N p- b d N q) =\nonumber\\
&=&(a d-b c)(N p-2 r q)=(N p-2 r q)=n.
\end{eqnarray}
In other words, the parameter $n$ is invariant under Morita equivalence.
The transformations of the  parameter $m$ is more subtle, since
its definition involves two new integers $(k,s)$, which are
implicitly defined by the Diophantine equation
\begin{equation}
(2 r) s- k N=1.
\end{equation}
By imposing that $s^\prime$ and $k^\prime$ satisfy the same equation with
$N^\prime$ and $2 r^\prime$, we obtain
\begin{equation}
s^\prime= c k +d s\ \ \  k^\prime= a k+ b s.
\end{equation}
Therefore, we have
\begin{eqnarray}
m^\prime&=& k^\prime q^\prime - p^\prime s^\prime=
(a k+ b s)(c p + d q)-(c k+ d s)(a p+b q)=\nonumber\\
&=&(a d-b c)( k q - p s)=k q-p s=m.
\end{eqnarray}
Namely, also the parameter $m$ is invariant under the Morita equivalence.  A straightforward
computation shows that also the last one is unaffected. Thus our model naturally  encodes this
symmetry, all his constituents being unchanged.

\section{The relation with the noncommutative Wilson lattice action}
\noindent
In Sect.2 we have tried to realize general projective
modules at discretized level: the approach we have followed is similar, in the spirit,
to the operatorial construction adopted by Connes and Rieffel in their seminal paper \cite{cori}
on continuous noncommutative YM$_2$.
Alternatively, as thoroughly reviewed by Konechny and Schwarz in \cite{KS}, these modules
can be also understood in terms of algebra of functions endowed with deformed products: the classical
example is the Moyal product on Schwartz spaces. This second approach can be also adapted to the discrete
formulation with the advantage that a "formal" space-time lattice structure will show up. However we
have to stress that the emerging lattice will be only indirectly correlated to the original one: it will posses
a different number of sites and a different $\theta$. Technically we will obtain a discretized
representation of the torus generated by the translations\footnote{This point has been often overlooked
in the literature. Also in the continuum version of the theory when written in the star-product formalism
$\theta$  is not that one of the original torus describing instead the derivations algebra. Morita
equivalence, nevertheless, connects these two tori.}.

\noindent Writing our reduced TEK model as a lattice-like
Wilson action proceeds through a by now well-established series of
steps \cite{s,sm}. However, some attention has to be paid on
initial set of geometrical data to be employed.  The $\theta$
associated with our reduced TEK model is $m/n$ and as in Sect. 2
the consistency of the lattice construction would force $m$ to be
even. Actually, this is not a real limitation: in fact if we fix
the geometrical data $N,\ 2r,\ p$ and $q$, the ambiguity intrinsic
in the Diophantine equation  determining  $m$ allows us to choose
it even.

\noindent
With this remark in mind one begins by defining the operator $\Delta(\vec{x})$ which realizes the
mapping between operators and functions on the lattice. In terms of twist-eaters
\begin{eqnarray}
Z_1&=&(\Gamma_1)^{kq-sp},\nonumber\\
Z_2&=&\Gamma_2,
\end{eqnarray}
we have ($\displaystyle{\vec{k}=\Bigl(2\pi\frac{m_1}{na},2\pi\frac{m_2}{na}\Bigr)}$)
\begin{equation}
\hat\Delta(\vec{x})=\frac{1}{n^2}\sum_{m_{1,2}}\exp(i\vec{k}\cdot\vec{x})\exp(i\pi\frac{m\,m_1m_2}{n})Z^{m_1}_1Z^{m_2}_2.
\end{equation}
The vector  $\vec{x}$ "corresponds" to the point $(n_1a , n_2 a)$ on a lattice of size $n$ (as usual we
denote $a$ as the lattice spacing). Then to each operator $\hat{f}$ (matrix in our discretized case) we can associate
the function
\begin{equation}
f(\vec{x})=n{\rm Tr}\Bigl[\hat{f}\hat\Delta(\vec{x})\Bigr]
\end{equation}
and conversely to each function $f(\vec{x})$ on the lattice
\begin{equation}
\hat{f}=\sum_{\vec{x}}f(\vec{x})\hat\Delta(\vec{x}).
\end{equation}
That this mapping be an isomorphism of algebras naturally indicates how to deform the product between
functions and in details we have
\begin{eqnarray}
f(\x)\star g(\x)&=&{n} {\rm Tr}\Bigl[\hat g \hat f \hat \Delta(\x)\Bigr]=n
\sum_{\y,\vec{z}}f(\vec{z}) g(\y) {\rm
Tr}\Bigl[\hat\Delta(\y)\hat\Delta(\vec{z})\hat
\Delta(\x)\Bigr]=\nonumber\\
&=&\sum_{\y,\vec{z}}f(\vec{z}) g(\y){\cal K}(\x-\y,\x-\vec{z}).
\label{starpro}
\end{eqnarray}
\noindent
The definition of $\mathcal{K}(\x-\y,\x-\vec{z})$ is given by
\begin{eqnarray}
\label{kernelq}
{\cal K}(\x-\y,\x-\vec{z})&=&{n}{\rm
Tr}\Bigl[\hat\Delta(\y)\hat\Delta(\vec{z})\hat\Delta(\x)\Bigr]
\end{eqnarray}
and in the following we shall compute the rhs. To reduce the amount of algebra is
useful to define the combination
\begin{equation}
u_{\vec{k}}=\exp\left(-i\pi \frac{m m_1 m_2}{n}\right) Z_1^{m_1} Z_2^{m_2},
\end{equation}
which satisfies the simple composition rule
\begin{equation}
u_{\vec{k}} u_{\vec{\ell}}=\exp(\pi i\Theta(k_1 \ell_2- k_2\ell_1 ))u_{\vec{k}+\vec{l}}
\end{equation}
and the trace property
\begin{equation}
\mathrm{Tr}(u_{\vec{k}} u_{\vec{\ell}})= n \delta_{\vec{k}+\vec{l}},
\end{equation}
with $\displaystyle{\Theta=m/n}$. Now the rhs of eq.
(\ref{kernelq}) can be rewritten as
\begin{eqnarray}
&=&\frac{1}{n^5}\sum_{\kk,\kk^\prime, \vec{\ell}}
e^{i\kk\cdot \x +i\kk^\prime\cdot\y+i\vec{\ell}\cdot\vec{z}}{\rm
Tr}({u}_{-\vec{k}^\prime} {u}_{-\vec{\ell}}\ {u }_{-\vec{k}}
)=\nonumber\\
&=&
 \frac{1}{n^5}\sum_{\kk,\kk^\prime,
\vec{\ell}} e^{i\kk\cdot \x
+i\kk^\prime\cdot\y+i\vec{\ell}\cdot\vec{z}}e^{\pi i \Theta
(k_1^\prime k_2+k_1^\prime\ell_2+k_2\ell_1-k_1
k_2^\prime-k^\prime_2\ell_1-k_1\ell_2)}{\rm
Tr}({u}_{-\vec{k}^\prime-\vec{\ell}-\vec{k}} )=\nonumber
\\
&=&
 \frac{1}{n^4}\sum_{\kk^\prime, \vec{\ell}} e^{
i\kk^\prime\cdot(\y-\x)+i\vec{\ell}\cdot(\vec{z}-\x)}e^{\pi i
\Theta (k_1^\prime \ell_2-k^\prime_2\ell_1)}=
\nonumber\\
&=&
 \frac{1}{n^4}\sum_{\kk^\prime, \vec{\ell}}
e^{i k^\prime_1(y_1-x_1+\pi \Theta \ell_2) +i
k^\prime_2(y_2-x_2-\pi \Theta
\ell_1)+i\vec{\ell}\cdot(\vec{z}-\x))}.
\end{eqnarray}
The sum over the momentum $\kk$ imposes that the two combinations,
\begin{equation}
\frac{y_1-x_1+\pi\Theta\ell_2}{a}=  \frac{1}{a}\left(m_1 a-n_1
a+\frac{2\pi^2}{n a} j_2 \frac{n m}{4\pi^2} a^2\right)=\left(m_1 -n_1+
\frac{m}{2} j_2\right)
\end{equation}
and
\begin{equation}
\frac{y_2-x_2-\pi\Theta\ell_1}{a}=\frac{1}{a}\left(m_2 a-n_2
a-\frac{2\pi^2}{n a} j_1 \frac{n m}{4\pi^2} a^2\right)=\left(m_2 -n_2-
\frac{m}{2} j_1\right),
\end{equation}
vanish modulo $n$. Notice that since $\displaystyle{\vec{j}= \frac{n a \vec{\ell}}{2\pi}}$ is an integer
vector, we must require that
\begin{equation}
j_1=\frac{2(n_2-m_2+h_1 n)}{m} \ \ \ {\rm and }\ \ \
j_2=\frac{2(m_1-n_1+h_2 n)}{m}
\end{equation}
belong to $\mathbb{Z}$ for a suitable choice of $\vec{h}$. Recall also that $m$ can be always
taken even as stressed before. These
conditions can be always met: in fact, the equations $ m j_i/2- n
h_i= (\mbox{a given integer})$ are always solvable when $n$ and
$m/2$ are coprime. Denoting, with $s^\prime$ and $k^\prime$ the
solution of the Diophantine equation
\begin{equation}
\label{B14}
{s^\prime m}-k^\prime n=1,
\end{equation}
the form of $j_i$ and $h_i$ is given by
\begin{equation}
j_1=2 s^\prime (n_2-m_2)\  \  \  \ h_1=k^\prime (n_2-m_2)\Rightarrow\
\ell_1=-\frac{4\pi s^\prime}{n a^2}(y_2-x_2)
\end{equation}
and
\begin{equation}
j_2=2 s^\prime (m_1-n_1)\  \  \  \ h_1=k^\prime (m_1-n_1)\Rightarrow\
\ell_2=\frac{4\pi s^\prime}{n a^2}(y_1-x_1).
\end{equation}
Substituting this result in the sum we finally have
\begin{eqnarray}
&&{\cal K}(\y-\x,\vec{z}-\x)=
\frac{1}{n^2}\exp\frac{4\pi i
s^\prime}{na^2}\Bigl[(y_1-x_1)(z_2-x_2)-(y_2-x_2)(z_1-x_1)\Bigr].
\label{kernel}
\end{eqnarray}
We can easily see how the star-product is explicitly realized: let us consider
the basis for the functions on the toroidal lattice given by the
$u_{\kk}(\x)=\exp(i\vec{k}\cdot\x)$ as defined in Sect.2. Their algebra is
now modified by the star product:
\begin{eqnarray}
u_{\kk}(\x)\star u_{\kk^\prime}(\x)&=&\sum_{\y,\vec{z}}
u_{\kk}(\vec{z}) u_{\kk^\prime}(\y){\cal K}(\x-\y,\x-\vec{z})=\nonumber\\
&=&\frac{1}{n^2}\sum_{\y,\vec{z}} e^{\frac{4\pi i
s^\prime}{n a^2}[(y_1-x_1)(z_2-x_2)-(y_2-x_2)(z_1-x_1)]
+i\kk\cdot\vec{z}+i\kk^\prime\cdot\y}=\nonumber\\
&=&\frac{1}{n^2}\sum_{\y,\vec{z}} e^{i y_1 [\frac{4\pi
s^\prime}{n a^2}(z_2-x_2)+k_1^\prime ]-i y_2 [\frac{4\pi
s^\prime}{n a^2}(z_1-x_1)-k_2^\prime]+ \frac{4\pi i s^\prime}{n a^2} (z_1 x_2-z_2
x_1)+i\kk\cdot \vec{z}}.
\end{eqnarray}
To evaluate this expression, we note that the sum over $\vec{y}$ is zero unless the following combinations
vanish
\begin{equation}
\frac{4\pi  s^\prime}{n a^2}(z_2-x_2)+k_1^\prime=\frac{2\pi}{n a}( 2 s^\prime
(m_2-n_2)+ j_1 )=0
\end{equation}
and
\begin{equation}
\frac{4\pi  s^\prime}{n a^2}(z_1-x_1)-k^\prime_2=\frac{2\pi}{n a}( 2 s^\prime
(m_1-n_1) - j_2 )=0.
\end{equation}
Here we have set $\x=\vec{n} a$, $\vec{z}=\vec{m}a$ and
$\kk^\prime= 2\pi\vec{j}/n a$. Taking into account  the
periodicity in $n$, the integer solutions  of the above equation
are the solutions of the Diophantine equations
\begin{equation}
2 (m_2-n_2) s^\prime -h_1 n=-j_1 \ \ \ \ {\rm and}\ \ \ \ 2 (m_1-n_1) s^\prime
-h_2 N=j_2.
\end{equation}
Comparing  this equation with eq. (\ref{B14}), we obtain the
solutions
\begin{eqnarray}
&&m_2=-j_1 \frac{m}{2}+n_2\ \Rightarrow\ z_2=x_2- \frac{n a^2 m}{4\pi}
k_1^\prime\nonumber\\
&&h_1=-k j_1\nonumber\\
&&m_1=j_2 \frac{m}{2}+n_1\ \Rightarrow\ z_1=x_1+  \frac{n a^2 m}{4\pi}
k_2^\prime\nonumber\\
&&h_2=k j_2.
\end{eqnarray}
Then the product of two Block-waves is
\begin{eqnarray}
&=& e^{\frac{4\pi i s^\prime}{n a^2} \left(\left(x_1+\frac{n a^2 m}{4\pi}
k_2^\prime\right) x_2-\left(x_2-\frac{n a^2 m}{4\pi}
k_1^\prime\right) x_1\right)+i k_1\left(x_1+\frac{n a^2 m}{4\pi}
k_2^\prime\right)+ik_2 \left(x_2-  \frac{n a^2 m}{4\pi}
k_1^\prime\right) }=\nonumber\\
&=& e^{i(\vec{k}+\kk^\prime)\cdot\vec{x}+\pi i \left(\frac{n a^2
m}{4\pi^2}\right)(k_1 k_2^\prime-k_2 k_1^\prime)}= e^{\pi i\Theta
(k_1 k_2^\prime-k_2 k_1^\prime)} u_{\kk+\kk^\prime}(\x).
\end{eqnarray}
\noindent
Having obtained an explicit discretized form of the star-product, we can construct a Wilson-like action:
following \cite{sm}, it is obtained through some algebraic manipulations. We rewrite
\begin{eqnarray}
S_{TEK}=\frac{\beta N}{n^2}\sum_{\vec{x}}{\rm Tr}\Bigl[&(&\exp(\pi i m/n)V_1 V_2-\exp(-\pi i m/n)V_2 V_1)\Bigr.\nonumber\\
\Bigl.&(&\exp(-\pi i m/n)V_2^\dagger V_1^\dagger-\exp(\pi i m/n)V_1^\dagger
V_2^\dagger)\hat\Delta(\vec{x})\Bigr],
\end{eqnarray}
using the new variables $V_i=\hat{V}_i\Gamma_i$ as
\begin{eqnarray}
S_{TEK}&=&2\beta Nn-\beta \frac{N}{n}\sum_{x}\sum_{i\neq j}
\hat{\cal V}_i(\vec{x})\star\hat{\cal V}_j(\vec{x}+a\vec{i})\star
\hat{\cal V}^*_i(\vec{x}+a\vec{j})\star\hat{\cal V}^*_j(\vec{x})=\nonumber\\
&=&\frac{1}{p-\theta q}\left(2\beta n^2-\beta \sum_{x}\sum_{i\neq j}
\hat{\cal V}_i(\vec{x})\star\hat{\cal V}_j(\vec{x}+a\vec{i})\star
\hat{\cal V}^*_i(\vec{x}+a\vec{j})\star\hat{\cal V}^*_j(\vec{x})\right)
\end{eqnarray}
where the explicit form of the discretized star-product eqs. (\ref{starpro}) and (\ref{kernel}) has been taken
into account. This is the promised Wilson-like representation of our TEK model,
 improved by a zero-point subtraction, and expressed just in terms of
geometrical data $(p,q)$, $\theta$, $n$ and $a$.
We must stress few facts: first of all
the dimensions of the lattice is $n^2$ and not $N^2$. This is expected:
in fact, as noted in \cite{sm}, the dimension of the lattice
is related to the representation of the translation operators. Accordingly
the star-product realizes a torus with $\theta=m/n$ and not $\theta=2r/N$, which is
the noncommutative parameter of a derivative torus: this again is in complete agreement with \cite{sm}, where the
Morita dual parameter appears.

\end{document}